\documentclass[journal]{IEEEtran}
\usepackage{mathrsfs}
\usepackage{graphicx}
\usepackage{url}
\usepackage{amsmath,amssymb,amsfonts}

\usepackage{epstopdf}
\usepackage{multirow}
\usepackage{ulem}
\usepackage[labelsep=period]{caption}

\usepackage{listings,mdframed,xcolor}

\definecolor{codeBackground}{rgb}{0.98, 0.8, 0.91}
\lstnewenvironment{mylisting}{%
   \lstset{
  rulecolor=\color{black},
  rulesepcolor=\color{gray},
  language=JAVA,
  aboveskip=3mm,
  belowskip=3mm,
  showstringspaces=false,
  columns=flexible,
  basicstyle=\footnotesize\ttfamily,
  numbers=none,
  numberstyle=\tiny\color{gray},
  keywordstyle=\color{blue},
  commentstyle=\color{dkgreen},
  stringstyle=\color{mauve},
  breaklines=true,
  breakatwhitespace=true,
  tabsize=3
   }%
   \mdframed[backgroundcolor=codeBackground,shadow=true,shadowsize=2pt,shadowcolor=black!30]%
}{%
   \endmdframed\ignorespaces
}

\let\emph\textit

\usepackage{color}
\usepackage{array}
\definecolor{dkgreen}{rgb}{0,0.6,0}
\definecolor{gray}{rgb}{0.5,0.5,0.5}
\definecolor{mauve}{rgb}{0.58,0,0.82}
\newcolumntype{P}[1]{>{\centering\arraybackslash}p{#1}}
\usepackage{hyperref}
\usepackage{pifont}
\newcommand{\xmark}{\ding{55}}

\def\ie{\textit{i.e.}}
\def\etal{\textit{et al.}}

\def\eg{\textit{e.g.}}

\ifodd 0
\newcommand{\HL}[1]{\textcolor{blue}{{#1}}}			   %COMMENTS BY Henry
\else
\newcommand{\HL}[1]{#1}
\fi

\usepackage[english]{babel}
\begin{document}
%title{The smart contract: A survey}
%
%
%\author[1]{Zibin Zheng}
%\author[1]{Shaoan Xie}
%\author[2]{Hongning Dai}
%\author[4]{Xiangping Chen}
%\author[3]{Huaimin Wang}
%
%
%\affil[1]{School of Data and Computer Science, Sun Yat-sen University Guangzhou, China  }
%
%\affil[2]{Faculty of Information Technology, Macau University of Science and Technology, Macau, SAR   }
%
%\affil[3]{National Laboratory for Parallel \& Distributed Processing \protect\\
%National University of Defense Technology, Changsha 410073 China  }
% \affil[4]{Institute of Advanced Technology,National Engineering Research Center of Digital Life \protect\\ Sun Yat-sen University, Guangzhou, China}
%
%
%\affil[ ]{\textit {Email: zhzibin@mail.sysu.edu.cn }}

%\begin{frontmatter}

%\history{Date of publication xxxx 00, 0000, date of current version xxxx 00, 0000.}
%\doi{xx.xxxx/ACCESS.2018.DOI}

\title{An Overview on Smart Contracts: Challenges, Advances and Platforms}
%\author{Zibin Zheng\authorrefmark{1}, \IEEEmembership{Senior Member, IEEE},
%Shaoan Xie\authorrefmark{1}, Hong-Ning Dai\authorrefmark{2}, \IEEEmembership{Senior Member, IEEE}, Weili Chen\authorrefmark{1}, Xiangping Chen\authorrefmark{1}, Jian Weng\authorrefmark{3}
%}
%\address[1]{School of Data and Computer Science, Sun Yat-sen University Guangzhou, China (e-mail: zhzibin@mail.sysu.edu.cn)}
%\address[2]{Faculty of Information Technology, Macau University of Science and Technology, Macau, SAR (e-mail: hndai@ieee.org)
% }
% \address[3]{Faculty of Information Technology, Macau University of Science and Technology, Macau, SAR (e-mail: hndai@ieee.org)
% }

\author{Zibin Zheng,Shaoan Xie, Hong-Ning Dai, Weili Chen, Xiangping Chen, Jian Weng, Muhammad~Imran
\thanks{Z. Zheng, S. Xie, W. Chen, X. Chen is with School of Data and Computer Science, Sun Yat-sen University, China.}
\thanks{H.-N. Dai is with Faculty of Information Technology, Macau University of Science and Technology, Macau (email: hndai@ieee.org).}
\thanks{J. Weng is with Jinan University, Guangzhou, China.}
\thanks{M. Immran is with College of Applied Computer Science, King Saud University, Riyadh, Saudi Arabia}
}

% \author[label1]{Zibin~Zheng}
% \address[label1]{School of Data and Computer Science, Sun Yat-sen University, China}

% \author[label1]{Shaoan Xie}
% %\address[label1]{School of Data and Computer Science, Sun Yat-sen University, China}

% \author[label2]{Hong-Ning~Dai}
% \address[label2]{Faculty of Information Technology, Macau University of Science and Technology, Macau SAR}

% \author[label1]{Weili~Chen}
% %\address[label1]{School of Data and Computer Science, Sun Yat-sen University, China}

% \author[label1]{Xiangping~Chen}
% %\address[label1]{School of Data and Computer Science, Sun Yat-sen University, China}

% \author[label3]{Jian~Weng}
% \address[label3]{College of Information Science and Technology, Jinan University, Guangzhou, China}

% \author[label4]{Muhammad~Imran}
% \address[label4]{College of Applied Computer Science, King Saud University, Riyadh, Saudi Arabia}
%\markboth
%{Zheng \headeretal: An Overview on Smart Contact: Platforms, Applications and Advances}
%{Zheng \headeretal: An Overview on Smart Contact: Platforms, Applications and Advances}
%
%\corresp{Corresponding author: Hong-Ning Dai (e-mail: hndai@ieee.org).}

\maketitle

\begin{abstract}
Smart contract technology is reshaping conventional industry and business processes. \HL{Being embedded} in blockchains, smart contracts enable the contractual terms of an agreement to be enforced automatically without the intervention of a trusted third party. As a result, smart contracts can cut down administration and \HL{save} services costs, improve the efficiency of business processes and reduce the risks. Although smart contracts are promising to drive the new wave of innovation in business processes, there are a number of challenges to be tackled. This paper presents a survey on smart contracts. We first introduce blockchains and smart contracts. We then present the challenges in smart contracts as well as recent technical advances. We also compare typical {smart contract}  platforms and give a categorization of {smart contract}  applications along with some representative examples.

\end{abstract}


%\end{frontmatter}

%\titlepgskip=-15pt

\section{Introduction}

Blockchain technology has recently fueled \HL{extensive} interests from both academia and industry. A blockchain is a distributed software system allowing transactions to be processed without the necessity of a trusted third party. As a result, business activities can be completed in an inexpensive and quick manner. Moreover, the immutability of blockchains also assures the distributed trust since it is nearly impossible to tamper any transactions stored in blockchains and all the historical transactions are auditable and traceable.

Blockchain technology is enabling \textit{smart contracts} that were first proposed in 1990s by Nick Szabo \cite{szabo1997idea}. In a smart contract, contract clauses written in computer programs will be automatically executed when {predefined}  conditions are met. Smart contracts consisting of transactions are essentially stored, replicated and updated in distributed blockchains. In contrast, conventional contracts need to be completed by a trusted third party in a centralized manner consequently resulting in long execution time and extra cost. The integration of blockchain technology with smart contracts will make the dream of a ``\textit{peer-to-peer market}'' come true.

\begin{figure}[h]
 \centering
 \includegraphics[width=8.6cm]{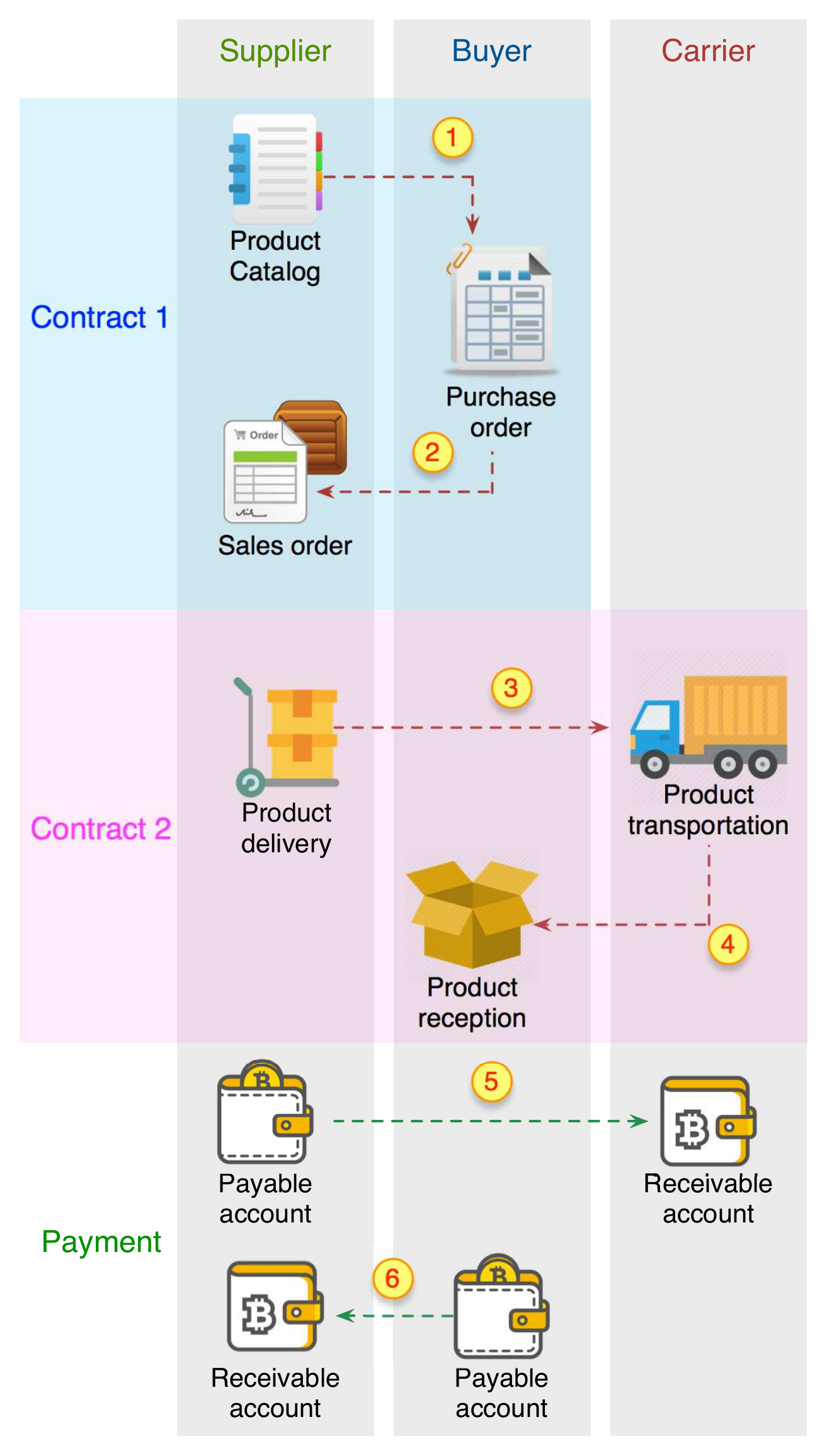}
 \caption{An example of a smart contract between a buyer and a supplier.}
\label{fig:smartcontract}
\end{figure}

Take a smart contract between a buyer and a supplier as an example. As shown in {Figure} \ref{fig:smartcontract}, a supplier first sends a product catalog to a buyer through the blockchain network. This catalog that includes product descriptions (such as property, quantity, price and availability) along with shipping and payment terms is stored and distributed in the blockchain so that a buyer can obtain the product information and verify the authenticity and reputation of the supplier at the same time. The buyer then submits the order with the specified quantity and payment date via the blockchain. This whole procedure forms a purchase contract (\ie, \emph{Contract 1}) enclosed in the blue box as shown in {Figure}~\ref{fig:smartcontract}. It is worth mentioning that the whole procedure is completed between the buyer and the supplier without the intervention of \HL{a} third party.

After \emph{Contract 1} is done, the supplier will search for a carrier in the blockchain to complete the shipping phase. Like \emph{Contract 1}, the carrier also publishes \HL{the} shipping description (such as transportation fees, source, destination, capacity and shipping time) as well as shipping conditions and terms in the blockchain. If the supplier accepts the contract issued by the carrier, the products will be delivered to the carrier who will finally dispatch the products to the buyer. This whole procedure constructs \emph{Contract 2} \HL{(enclosed in the pink box)} as shown in {Figure}~\ref{fig:smartcontract} . Similarly, the whole procedure of \emph{Contract 2} is also conducted without the intervention of a third party.

In addition to \HL{automatic execution}  of \emph{Contract 1} and \emph{Contract 2}, the payment procedures (including the payment from the supplier to the carrier and that from the buyer to the supplier) are also completed automatically. For example, once the buyer confirms the reception of the products, the payment between the buyer and the supplier will be automatically triggered as the predefined condition is met. The financial settlement from the buyer to the supplier is conducted via {crypto currencies}  (\HL{\eg,} Bitcoin or Ether\footnote{Commonly used acronyms in this paper are listed in Table \ref{tab:acronym}}.). In contrast to conventional transactions, the whole process is done in a peer-to-peer manner without the intervention of \HL{third parties} like banks. As a result, the turnaround time and transactional cost can be greatly saved.

In summary, smart contracts have the following advantages compared with conventional contracts:
\begin{itemize}
\item \textit{Reducing risks.} Due to the immutability of blockchains, smart contracts cannot be arbitrarily altered once they are issued. Moreover, all the transactions that are stored and duplicated throughout the whole distributed \HL{blockchain} system are traceable and auditable. As a result, malicious behaviors like financial frauds can be greatly mitigated.
\item \textit{Cutting down administration and service costs.} Block-chains assure the trust of the whole system by distributed consensus mechanisms without going through a central broker or a mediator. Smart contracts stored in blockchains can be automatically triggered in a decentralized way. Consequently, the administration and services costs due to the intervention from the third party can be \HL{significantly} saved.
\item \textit{Improving the efficiency of business processes.} The elimination of the dependence on the intermediary can significantly improve the efficiency of business process. Take the aforementioned supply-chain procedure as an example. The financial settlement will be automatically completed in a peer-to-peer manner once the {predefined}  condition is met (\eg, the buyer confirms the reception of the products). As a result, the turnaround time can be significantly reduced.
\end{itemize}

\begin{table}[t]
\centering
\caption{Acronym Table}
\renewcommand{\arraystretch}{1.5}
\label{tab:acronym}
\begin{tabular}{|l|l|}\hline 
{\textbf{Terms}}&{\textbf{Acronyms}}\\
\hline  \hline
{Proof of Work}&{PoW}\\
\hline
{Proof of Stake}&{PoS}\\
\hline 
{Practical Byzantine-Fault Tolerance}&{PBFT}\\
\hline 
{Low Level Virtual Machine}&{LLVM}\\
\hline 
{Convolutional Neural Network }&{CNN}\\
\hline 
{Long Short Term Memory}&{LSTM}\\
\hline 
{Ether}&{ETH}\\
\hline 
{Bitcoin}&{BTC}\\
\hline 
{Ethereum Virtual Machine}&{EVM}\\
\hline 
{Unspent Transaction Output}&{UTXO}\\
\hline 
{Internet of Things}&{IoT}\\
\hline 
{Distributed Autonomous Corporation}&{DAC}\\
\hline 
{Certificate Authority}&{CA}\\
\hline 
{Delegated Proof of Stake}&{DPOS}\\
\hline 
{WebAssembly}&{Wasm}\\
\hline 
{Border Gateway Protocol}&{BGP}\\
\hline 
\end{tabular}
\end{table}

Smart contracts are boosting a broad spectrum of applications ranging from industrial Internet of Things to financial services ~\cite{hndai:blockchain-iot2019,bogner2016decentralised,zhang2015iot,mccorry2017smart,luu2017smart,hillbom2016applications,yasin2016online,scoca2017smart,wan2019blockchain,moin2019securing}. Although smart contracts have great potentials to reshape conventional business procedures, there are a number of challenges to be solved. For example, even if blockchains can assure a certain anonymity of the parties of the contract, the privacy of the whole contract execution may not be preserved since all the transactions are globally available. Moreover, it is challenging to ensure the correctness of smart contracts due to \HL{vulnerabilities of computer programs to the faults and failures}.

\begin{table*}
\caption{{Comparison with related work}}
\label{tab:compwork}
\centering
\begin{tabular}{P{1.8cm}|P{1.6cm}|P{1.6cm}|P{2.2cm}|P{2.5cm}|P{1.6cm}|P{1.6cm}|P{1.6cm}}
\hline
{Research}&{Ethereum} & {Other platforms}&{Programming Languages}&{Other technical challenges}&{Technical advances} & {Rising challenges}& {Applications}\\
\hline\hline
\cite{zheng2017overview,omohundro2014cryptocurrencies,li2017survey}&\checkmark&\HL{\xmark}&\HL{\xmark}&\HL{\xmark}&\HL{\xmark}&\HL{\xmark}&\HL{\xmark}\\
\cite{atzei2017survey,delmolino2016step}&\checkmark&\HL{\xmark}&\checkmark&\HL{\xmark}&\HL{\xmark}&\HL{\xmark}&\HL{\xmark}\\
\cite{harz2018towards}&\checkmark&\checkmark&\checkmark&\HL{\xmark}&\HL{\xmark}&\HL{\xmark}&\HL{\xmark}\\
\cite{bartoletti2017empirical}&\checkmark&\checkmark&\checkmark&\HL{\xmark}&\HL{\xmark}&\HL{\xmark}&\checkmark\\
\cite{alharby2017blockchain,macrinici2018smart,wang2018overview} &\checkmark&\checkmark&\checkmark&\checkmark&\checkmark&\HL{\xmark}&\checkmark\\
This paper&\checkmark&\checkmark&\checkmark&\checkmark&\checkmark&\checkmark&\checkmark\\
\hline
\end{tabular}
\end{table*}

There are some recent studies on smart contracts. For example, \cite{zheng2017overview,omohundro2014cryptocurrencies,li2017survey} present comprehensive surveys of blockchain technology and \HL{briefly} introduce smart contracts. The work of \cite{atzei2017survey} provides an in-depth survey on Ethereum smart contract programming vulnerabilities while \cite{harz2018towards} presents a detailed survey over verification methods on smart contract languages. The work of \cite{delmolino2016step} \HL{reports} authors' experiences in teaching smart contract programming and summarizes several typical \HL{types} of mistakes \HL{made by students}. Ref. \cite{bartoletti2017empirical} \HL{presents} an empirical analysis on smart contract platforms. Recent studies \cite{alharby2017blockchain,macrinici2018smart} \HL{also collect some literature of smart contracts and present reviews while fail to discuss the challenges in this area}. \HL{Moreover, the} work of \cite{wang2018overview} presents a brief overview of smart contract platforms and architectures. \HL{However, most of existing papers fail to identify the rising challenges and give a comprehensive survey}. \HL{For example, Ethereum can be used to conduct illegal business such as Ponzi schemes that were reported to defraud over 410,000 US dollars while few studies address this issue~\cite{chen2018detecting}}. \HL{We summarize the differences between this paper and existing studies in Table \ref{tab:compwork}}.

%Most of existing studies are focusing on programming challenges of smart contracts but there are still many significant challenges to be tackled. For instance, detecting scam contracts is of great importance in the smart contract community.

The objective of this paper is to \HL{conduct} a systematic overview of technical challenges \HL{in smart contracts} enabled by blockchain technologies.  Contributions of \HL{this paper are highlighted} as following:
{
\begin{itemize}
\item {Important research challenges in the life cycle of smart contracts are identified.}
\item{Recent advances in addressing technical \HL{challenges} are summarized.}
\item{A detailed comparison of typical smart contract platforms is made.}
\item{Diverse smart contract applications are summarized.}
\end{itemize}}

\textbf{Organization of this paper}. Section \ref{concept} gives a brief introduction to blockchains and smart contracts. Section \ref{pra} then summarizes research challenges in smart contracts as well as recent technical advances. Section \ref{platforms} next compares typical smart contract \HL{development} platforms. Section \ref{application} categorizes typical smart contract applications. Finally, Section \ref{conclusion} concludes the paper.

\section{Overview of Blockchain and Smart Contract}
\label{concept}
Smart contracts are built upon blockchain technology \HL{ensuring} the correct execution of the contracts. We first provide a brief introduction to blockchain technology in Section \ref{subsec:blockchain}. We then give an overview on smart contracts in Section \ref{subsec:smart-contract}.

\subsection{Blockchain}
\label{subsec:blockchain}

\begin{figure*}[t]
 \centering
 \includegraphics[width=14.8cm]{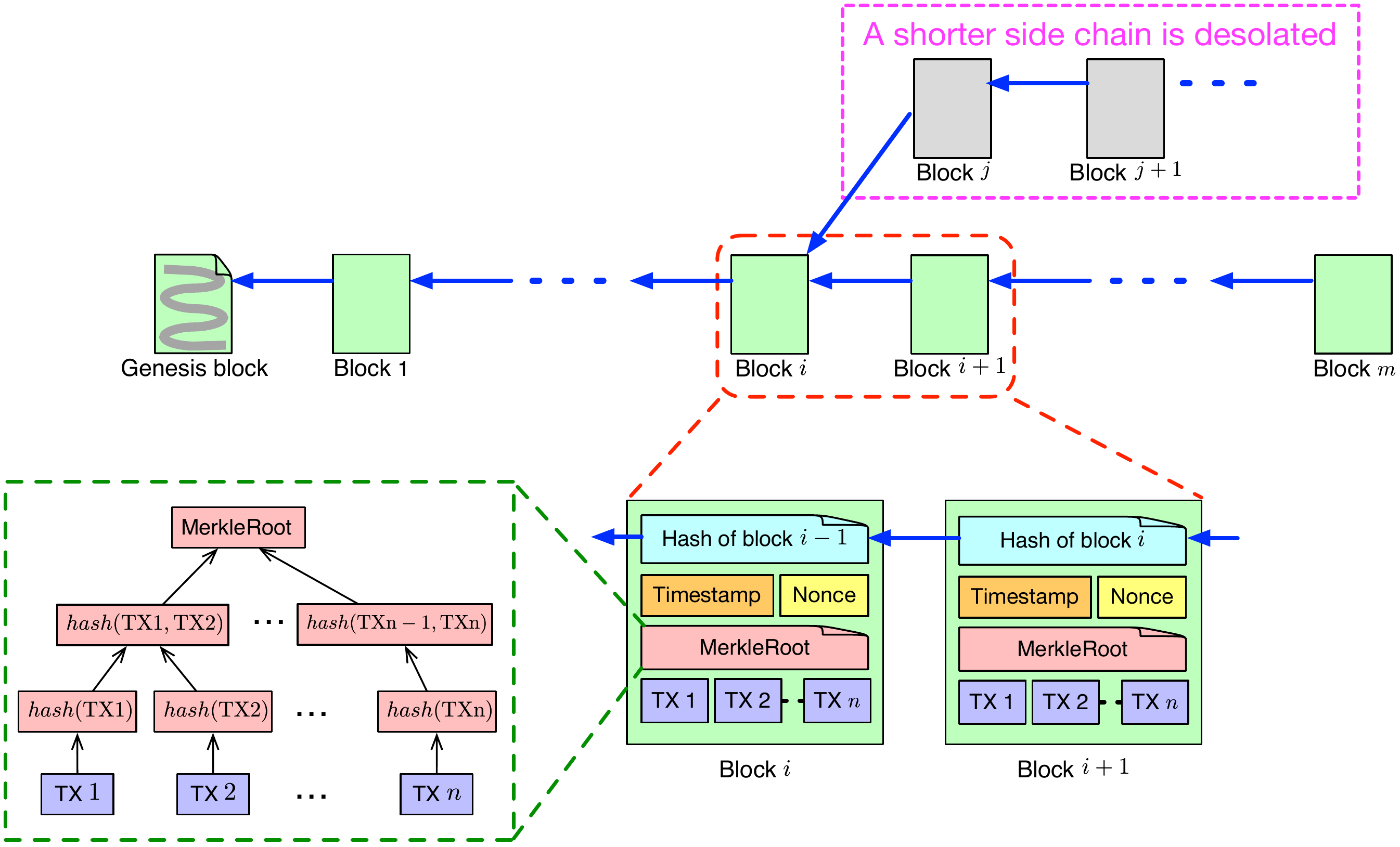}
 \caption{\HL{A blockchain consists} of a sequence of blocks, each of which contains \HL{an inverse} hash pointing back to its parent block. Meanwhile, there are a number of transactions stored inside a block.}
 \label{fig:blockchain}
\end{figure*}

A blockchain can be regarded as a public ledger, in which all transactions cannot be falsified. {Figure}   \ref{fig:blockchain} illustrates an example of a blockchain. A blockchain is a continuously-growing chain of blocks.  When a new block is generated, all the nodes in the network will participate in validating the block. Once the block is validated, it will be appended to the blockchain.

To validate the trustfulness of blocks, consensus algorithms are developed. Consensus algorithms determine which node to store the next block and how the new appended block to be validated by other nodes. Representative consensus algorithms include proof of work (PoW) \cite{nakamoto2008bitcoin} and proof of stake (PoS) and practical byzantine-fault tolerance (PBFT\HL{)} \cite{castro1999practical,king2012ppcoin}. Consensus algorithms are usually done by users who first solve the puzzle (\ie, PoW or PoS). These users are called \emph{miners}. {Each miner keeps a full copy of the blockchain}. Different from PoW and PoS, PBFT requires \HL{a} multi-round voting \HL{process} to reach the consensus. The distributed consensus algorithms can ensure that transactions are done without the intervention of third parties like banks. As a result, the transaction costs can be saved. Moreover, users transact with their virtual addresses instead of real identities so that the privacy of users is also preserved.

In blockchain systems, it is possible that several nodes can successfully reach the consensus (\ie, solving the puzzle) at the same time\HL{, consequently} it can cause the bisected branches. To solve the disparity, a shorted side chain is desolated as shown in {Figure} \ref{fig:blockchain} while the longest chain is selected as the valid chain. This mechanism is effective since the longer chain is more tolerant to malicious attacks than the shorter chain in distributed systems.

In summary, blockchain technology has the key characteristics of decentralization, immutable, persistency and anonymity \cite{zibin2016blockchain,tapscott2016blockchain,MLi:2018}.

\subsection{Smart Contract}
\label{subsec:smart-contract}

Smart contracts \HL{can be regarded as} a great advance \HL{in} blockchain technology \cite{ream2016upgrading}. In 1990s, \HL{a} smart contract \HL{was} proposed as a computerized transaction protocol that executes the contractual terms of an agreement \cite{szabo1997idea}. Contractual clauses that are embedded in smart contracts will be enforced automatically when a certain condition is satisfied (\eg, one party who breaches the contract will be punished automatically).

Blockchains are enabling smart contracts. Smart contracts are \HL{essentially} implemented on top of blockchains. The approved contractual clauses are converted into executable computer programs. The logical connections between  contractual clauses have also been preserved in the form of logical flows in programs (\eg, \HL{the} \texttt{if-else-if} statement). The execution of each contract statement is recorded as an immutable transaction stored in the blockchain. Smart contracts guarantee appropriate access control and contract enforcement. In particular, developers can assign access permission for each function in the contract. {Once any condition in a smart contract is} satisfied, the triggered statement will automatically execute the corresponding function in a predictable manner. For example, Alice and Bob agree on the penalty of violating the contract. If Bob breaches the contract, the corresponding penalty (as specified in the contract) will be automatically paid \HL{(deducted)} from Bob's deposit.

\begin{figure*}[t]
 \centering
 \includegraphics[width=15.6cm]{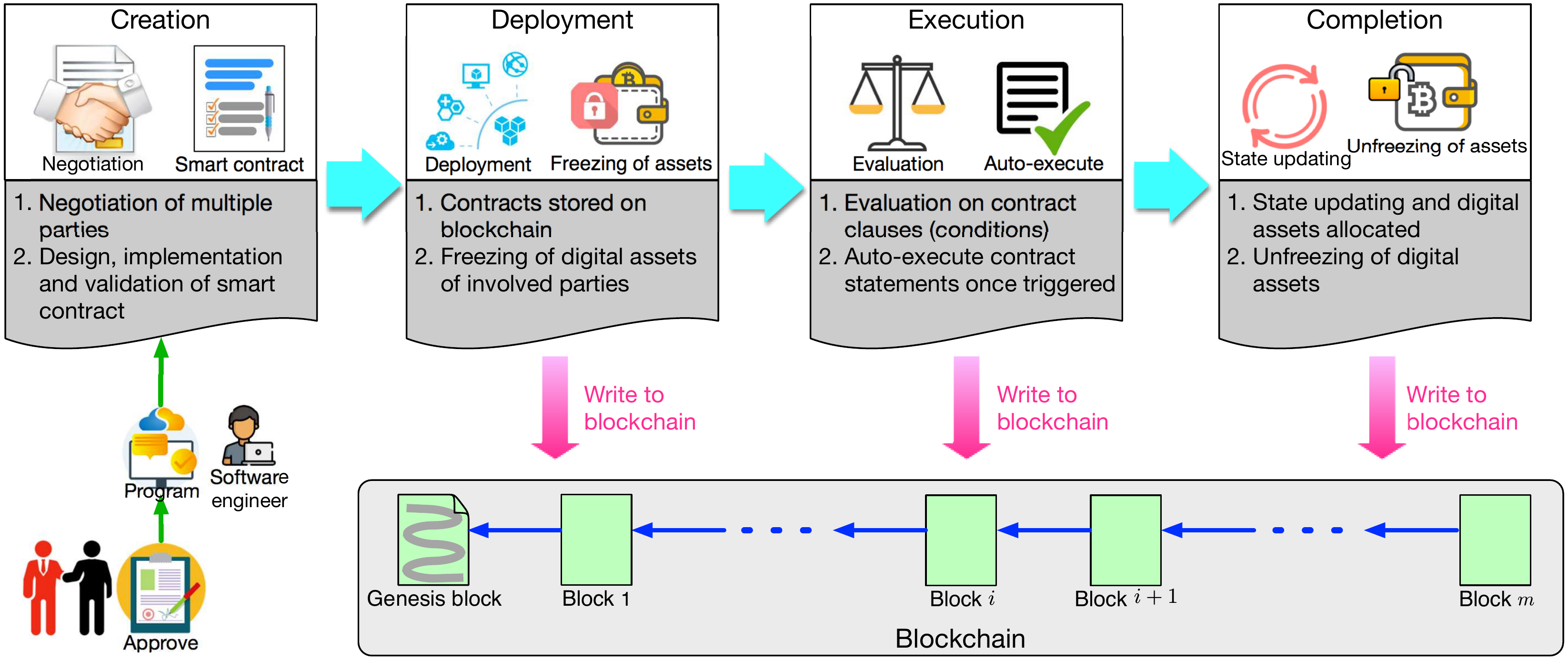}
 \caption{A life cycle of a smart contract consists of four major phases: Creation, Deployment, Execution and Completion}
 \label{fig:contract}
\end{figure*}

The whole life cycle of smart contracts consists of four consecutive phases as illustrated in {Figure}   \ref{fig:contract}:
\begin{enumerate}
\item  \textit{Creation} of smart contracts. Several involved parties first negotiate on the obligations, rights and prohibitions on contracts. After multiple rounds of discussions and negotiations, an agreement can reach. Lawyers or counselors will help parties to draft an initial contractual agreement. Software engineers then convert this agreement written in natural languages into a smart contract written in computer languages including declarative language\HL{s} and logic-based rule language\HL{s} \cite{Idelberger:2016}. Similar to the development of computer software, the procedure of the smart contract conversion is composed of design, implementation and validation (\ie, testing). It is worth mentioning that the creation of smart contracts is an iterative process involving with multiple rounds of negotiations and iterations. Meanwhile, it is also involved with multiple parties, such as stakeholders, lawyers and software engineers.

\item  \textit{Deployment} of smart contracts. The validated smart contracts can then be deployed to platforms on top of blockchains. Contracts stored on the blockchains cannot be modified due to the immutability of block-chains. Any emendation requires the creation of a new contract. Once \emph{smart} contracts are deployed on blockchains, all the parties can access the contracts through the blockchains. Moreover, digital assets of both involved parties in the smart contract are locked via freezing the corresponding digital wallets \cite{Sillaber2017}. For example, the coin transfers (either incoming or outgoing) on the wallets relevant to the contract are blocked. Meanwhile, the parties can be identified by their digital wallets.

\item  \textit{Execution} of smart contracts. After the deployment of smart contracts, the contractual clauses have been monitored and evaluated. Once the contractual conditions reach (\eg, product reception), the contractual procedures (or functions) will be automatically executed. It is worth noting that a smart contract \HL{consists} of a number of declarative statements with logical connections. When a condition is triggered, the corresponding statement will be automatically executed, consequently a transaction being executed and validated by miners in the blockchains \cite{koulu2016blockchains}. The committed transactions and the updated states have been stored on the blockchains thereafter.

\item  \textit{Completion} of smart contracts. After a smart contract has been executed, new states of all involved parties are updated. Accordingly, the transactions during the execution of the smart contracts as well as the updated states are stored in blockchains. Meanwhile, the digital assets have been transferred from one party to another party (\eg, money transfer from the buyer to the supplier). Consequently, digital assets of involved parties have been unlocked. The smart contract \HL{then} has completed the whole life cycle.
\end{enumerate}

It is worth mentioning that during deployment, execution and completion of a smart contract, a sequence of transactions has been executed (each corresponding to a statement in the smart contract) and stored in the blockchain. Therefore, all these three phases need to write data to the blockchain as shown in {Figure}~\ref{fig:contract}.

\section{Challenges and advances {of smart contract}}
\label{pra}

Although a smart contract is a promising technology, there are still a number of challenges to be tackled. We categorize these major challenges into four types according to four phases of the \HL{life cycle of smart contracts}. Meanwhile, we also give an overview on recent advances in solving these challenges. Table \ref{tab:challenges} summarizes the challenges and recent advances.

\begin{table*}[t]
\centering
\caption{Summary of challenges and advances in smart contracts}
\label{tab:challenges}
\renewcommand{\arraystretch}{1.5}
\footnotesize
\begin{tabular}{m{1.5cm}|m{5.5cm}|m{8.3cm}}
\hline
 \textbf{Phases}& \textbf{Challenges} & \textbf{Advances}\\
\hline\hline
\multirow{6}{*}{Creation} & & $\bullet$  Recover source code \cite{zhou2018erays}  \\
&  1) Readability  &  $\bullet$ Human readable code \cite{frantz2016institutions}, \cite{ciatto2018blockchain} \\
& &  $\bullet$ Human readable execution \cite{kasampalis2018iele}, \cite{lattner2004llvm} \\
\cline{2-3}
&  &  $\bullet$ Re-entrancy \cite{Coblenz:2017}, \cite{mavridou2018tool}, \cite{liu2018reguard} \\
& 2) Functional issues  &  $\bullet$ Block randness \cite{bonneau2015bitcoin}, \cite{lenstra2015random}, \cite{bunz2017proofs} \\
& &  $\bullet$ Overcharging \cite{chen2017under}, \cite{chen2018towards} \\
\hline

\multirow{6}{*}{Deployment} & & $\bullet$ Bytecode analysis
\cite{luu2016making}, \cite{albert2018ethir}, \cite{knecht2017smartdemap}, \cite{grech2018madmax}, \cite{krupp2018teether}, \cite{brent2018vandal}, \cite{grechgigahorse}, \cite{amani2018towards} \\
& 1) Contract correctness &  $\bullet$ Source code analysis \cite{Bhargavan:2016}, \cite{Swamy:2016}, \cite{kalra2018zeus}, \cite{mcmillan2007interpolants} \\
& &   $\bullet$  Machine learning based analysis \cite{liu2018s}, \cite{huang2018hunting}, \cite{tann2018towards}  \\
\cline{2-3}
&  & $\bullet$ Graph based analysis \cite{charlierprofiling}, \cite{frowis2017code} \\
& 2) Dynamic control flow &  $\bullet$ Path-searching  \cite{nikolic2018finding}\\
& & $\bullet$  Execution environment \cite{fu2019evmfuzz}\\
\hline

\multirow{6}{1.6cm}{Execution} & \multirow{2}{*}{1) Trustworthy oracle} & $\bullet$  Third-party involved \cite{zhang2016town} \\
&  &  $\bullet$ Decentralized \cite{adler2018astraea}, \cite{ellis2017decentralized}   \\
\cline{2-3}
&  \multirow{2}{*}{2) Transaction-ordering dependence}   &  $\bullet$ Sequential execution \cite{mavridou2017designing} \\
&  &  $\bullet$ Predefining contract  \cite{natoli2016blockchain} \\
\cline{2-3}
&  \multirow{2}{*}{3) Execution efficiency}   & $\bullet$ Execution serialization \cite{dickerson2017adding}, \cite{sergey2017concurrent}, \cite{anjana2018efficient} \\
& & $\bullet$ Inspection of contract  \cite{bragagnolo2018smartinspect}\\
\hline

\multirow{4}{*}{Completion} &  \multirow{2}{*}{1) Privacy and Security} & $\bullet$  Privacy \cite{kosba2016hawk}, \cite{zyskind2015enigma} \\
&  &   $\bullet$  Security  \cite{Apostolaki:2018}   \\
\cline{2-3}
&   \multirow{2}{*}{2) Scam}   &  $\bullet$ Ponzi scheme \cite{bartoletti2017dissecting} \cite{chen2018detecting}, \cite{bartoletti2018data} \\
&  &   $\bullet$  Honeypot  \cite{torres2019art} \\

\hline
\end{tabular}
\end{table*}

\subsection{Creation challenges}
\label{subsec:programming}

\emph{\HL{Contract creation}} is an important step to implement smart contracts. Users have to code their own contracts and then deploy them in various blockchain platforms (to be introduced in Section \ref{platforms}). Since blockchains are essentially immutable, blockchain-based smart contracts also cannot be modified after being deployed. As a result, developers need to carefully address the following problems. %In addition to the vulnerabilities listed by Atzei et al. \cite{atzei2017survey}, we also find that the \emph{gas-costly problem} also needs to be taken into account.

\subsubsection{Readability}

Most of smart contracts are written in programming languages such as Solidity, Go, Kotlin and Java (to be described in Section \ref{platforms}). Then source codes will be compiled and executed. Therefore, in different time periods, programs have different forms\HL{ of codes}. How to make programs readable in each form remains a big challenge. 
%However, the implementation of smart contracts in programming languages requires the programming capability of developers. How to make different forms still remains a bug challenge.
%As a result, it hinders the development of smart contracts since other domain experts like financial advisers and lawyers have to understand clauses written in programming languages \cite{delmolino2016step}. In addition, w
%How to improve the readability of smart contracts becomes a challenge.

\vspace*{0.1cm}
\hspace*{-0.25cm}\textbf{Recent advances {for readability challenge}}
\vspace*{0.1cm}

\begin{itemize}

\item \textit{Recover source code}: It is shown in \cite{zhou2018erays} that \HL{more than} 77\% smart contracts have not released public source codes, all of which are involved with over \$3 billion US dollars. \HL{Unavailability of} source code makes smart contract be opaque to the official auditors. To address this issue, \cite{zhou2018erays} proposed a reverse engineering tool (namely Erays) to analyze compiled smart contracts. This reverse engineering tool is able to convert hex encoded contract into a human readable pseudo codes.

\item \textit{Human readable code}: Frantz and Nowostawski \cite{frantz2016institutions} proposed a semi-automated translation system that can convert human-readable contract representations into computational programs. Essentially, this semi-automated translation system \HL{has been} implemented according to the concept from institutional analysis \cite{ostrom2014collective}. In particular, institution specifications can be decomposed into different components such as \texttt{attributes}, \texttt{deontic}, \texttt{aim}, \texttt{conditions} and \texttt{or else}. These components are then mapped into the corresponding blocks written in programming languages. For example, the \texttt{attributes} can be converted into \texttt{structs} in Solidity. \HL{Since} most of smart contract programming languages are object-oriented languages, \cite{ciatto2018blockchain} argues that declarative language backed by a logic programming computational model might be more suitable for smart contract. For example, \HL{the authors claim} that Prolog (a type of logic language) does not requires compilation so it also avoids the inspection on smart contract compilation.

\item \textit{Human readable execution}: Although many platforms attempt to provide smart contract developers with high level languages, these smart contracts will then be compiled into other forms, e.g., bytecode in \HL{\emph{Ethereum Virtual Machine} (EVM)}. In most of cases, two parties in the transaction need to understand the contracts at the level \HL{that} it has been stored and executed on blockchain. Ref. \cite{kasampalis2018iele} proposed an intermediate level language named IELE to solve this challenge. IELE has a similar syntax to Low Level Virtual Machine (LLVM) \cite{lattner2004llvm} so as to provide compilers with high-level information during \HL{the compile time, link time, run time, and idle time}.
%In spite of the advances in automating smart contracts, there is still a barrier between human-readable languages and programming languages used for smart contracts. It will be a future direction in developing smart contracts.
\end{itemize}

\subsubsection{Functional issues}
There are a number of functional issues with incumbent {smart contract}  platforms. We present several representative challenges: 1) \textit{Re-entrancy} means that the interrupted function can be safely recalled again. Malicious users may exploit this vulnerability to steal digital currency as indicated in~\cite{LI:FGCS2017}.  2) {\it Block randomness.} Some smart contract applications such as lotteries and betting pools may require randomness of generated blocks. This can be achieved by generating pseudo-random numbers in a block timestamp or nonce. However, some malicious miners may fabricate some blocks to deviate from the outcome of the pseudo-random generator. In this way, attackers can control the probability distribution of the outcomes as shown in \cite{Bonneau:eprint-2015}. 3) {\it Overcharging}. It is shown in recent work~\cite{chen2017under} that smart contracts can be overcharged due to the under-optimization of smart contracts. These overcharged patterns have the features like dead code, expensive operations in loop\HL{s consisting of repeated computations}.

\vspace*{0.1cm}
\hspace*{-0.25cm}\textbf{Recent advances {for functional issues}}
\vspace*{0.1cm}

\begin{itemize}

\item \textit{Re-entrancy}: Recently, several proposals attempt to solve some of the above challenges. Obsidian \cite{Coblenz:2017} was proposed to address re-entrancy attacks and money leakage problems. In particular, Obsidian exploits {\it named} states to enable consistency checking on state transitions and verification so that re-entrancy vulnerability can be mended. Moreover, a \HL{data flow} analytical method was proposed to prevent the illegal digital currency stealing from the leakage. Ref. \cite{mavridou2018tool} proposed to eliminate re-entrancy vulnerabilities by prohibiting the nesting calling among functions in the contract. Liu \etal~\cite{liu2018reguard} proposed to perform the fuzz testing on smart contracts by iteratively generating random but diverse transactions to detect re-entrancy bugs.

\item \textit{Block randness}: Blockchain is regarded as a promising technology to generate public and unpredictable random values. However, the random output might not be so random as people expect. Miners could control the block generation and release the block until they find it profitable. To address this issue, \cite{bonneau2015bitcoin} proposed to use the delay-function to generate randomness. It means that the random value \HL{will} be only be known to others after a short time period since its generation. In this way, the blockchain moves on and the miners could not withhold their blocks to profit. But delay functions are not suitable for smart contracts as most of they require instance verification. To this end, \cite{lenstra2015random} proposed the Sloth function to allow faster verification. Based on \cite{lenstra2015random}, \cite{bunz2017proofs} proposed \HL{a} multi-round protocol \HL{to verify} delay functions using a refereed delegation model. It reduces the cost of verifying the output from \$30 to \$0.4.

\item \textit{Overcharging}: Besides from caring the efficiency of their programs, developers of smart contract also need to pay attention to their execution costs. Ref. \cite{chen2017under} reported that over 90\% of real smart contracts suffer from gas-costly patterns in Ethereum. Chen \etal~\cite{chen2018towards} proposed GasReducer, a tool used to detect gas-costly patterns. \HL{GasReducer can} replace under-optimized bytecode with efficient bytecode.

\end{itemize}
%\textbf{Verification}

%\subsubsection{Verifiability}

%Most of incumbent {smart contract}  platforms (including Ethereum, Fabric and Rootstock) are Turing complete. Therefore, they can represent smart contracts more expressively than Corda, Stellar and Bitcoin, which are Turing incomplete. However, the Turing-completeness of smart contracts also results in potential software bugs as indicated in \cite{Dinh:TKDE2018}. For example, Decentralized Autonomous Organization (DAO) attack \cite{thedaoattack} was mainly caused by the recursive-calling vulnerability of Ethereum. It is reported in \cite{thedaoattack} that over 60 million dollars was stolen in the attack and Ethereum finally chose to hardfork the Ethereum blockchain to get those stolen ethers back.

%\vspace*{0.1cm}
%\hspace*{-0.25cm}\textbf{Recent advances}
%\vspace*{0.1cm}

%Recently, Pact\footnote{http://pact-language.readthedocs.io/en/latest/} was proposed to offer the solution to the verifiable problem. Pact is a declarative language supporting the atomic execution like conventional database management systems (DBMS), in which the whole operation will be rolled back if any failure occurs \cite{popejoy2016pact}. Corda and Stellar trade the expressiveness for the verifiability of smart contracts \cite{Dinh:TKDE2018}. In this way, the aforementioned security risks can be avoided.

\subsection{Deployment challenges}

After creation, smart contracts will be deployed on blockchain platforms. But smart contracts need to be checked carefully to avoid potential bugs. Furthermore, {smart contract}  developers need to be aware of the contract's interaction patterns to mitigate potential losses due to the malicious behaviors (such as frauds and attacks \cite{thedaoattack}). We next describe the challenges as well as advances in {smart contract}  deployment.

\subsubsection{Contract correctness}

Once smart contracts have been deployed on blockchains, it is nearly impossible to make any revisions. Therefore, it is of vital importance to evaluate the correctness of smart contracts before the formal deployment. However, it is challenging to verify the correctness of smart contracts due to the complexity of modelling \HL{smart contracts}.

\vspace*{0.1cm}
\hspace*{-0.25cm}\textbf{Recent advances {for contract correctness}}
\vspace*{0.1cm}

\begin{itemize}

\item \textit{Bytecode analysis}: Bytecode level analysis only requires the compiled bytecode\HL{s} of smart contracts, which are much easier to obtain. How to utilize these bytecode to detect security threats has become a hot research topic. In particular, OYENTE was proposed in \cite{luu2016making} to identify potential security bugs including mishandled exceptions and timestamp-\HL{dependent} problems. Based on the control graph generated by OYENTE, \cite{albert2018ethir} produces the rule-based representations for high level bytecode analysis. Meanwhile, Knecht and Stiller  \cite{knecht2017smartdemap} proposed a smart contract deployment and management platform (SmartDEMAP) to address the trust problem during the contract development and deployment. Moreover, other code quality control tools such as automated bug-finders can also be equipped with SmartDEMAP. In this manner, smart contracts can be deployed only after the trustful conditions are fulfilled. Ref. \cite{grech2018madmax} proposed MadMax to predict gas-focus vulnerabilities in Ethereum smart contracts. The combination of control-flow-analysis-based decompiler and declarative program-structure queries enables the method detecting the vulnerabilities in high precision. Meanwhile \cite{tsankov2018securify} symbolically analyzed the contract dependency graph to extract precise semantic information from the code. Then, it checks compliance and violation patterns that capture sufficient conditions to prove if a property holds or not. Furthermore, \cite{krupp2018teether} proposed \HL{a method} to search for certain critical paths in the control flow graph of a smart contract and identify paths that \HL{may} lead to a critical instruction, where the arguments of instructions can be controlled by an attacker. Ref. \cite{brent2018vandal} proposed Vandal, a tool which firstly converts low level bytecode into register transfer language \HL{that was then translated} into logic semantic relations. In addition, \cite{grechgigahorse} proposed Gigahorse, a tool which is able to decompile smart contract bytecode to high level 3-address code representation. The new intermediate representation of smart contracts makes \HL{the} implicit data and control flow dependencies of the EVM bytecode \HL{be} explicit. Amani \etal~\cite{amani2018towards} \HL{reconstructed} bytecode sequences into blocks of straight-line code and created a program logic to \HL{identify} the security \HL{vulnerability} of the contract.

\item \textit{Source code analysis}: Compared with bytecode level analysis, source code analysis requires the \HL{availability of smart contract source codes}. \HL{Although} the source code analysis contains more information, it \HL{also requires highly-}precise analysis. There are a number of studies on source code analysis of smart contracts. In particular, a formal verification method was proposed in \cite{Bhargavan:2016} to analyze and verify both the runtime safety and the functional correctness of smart contracts (\eg, Ethereum contracts). This method first translates smart contracts into codes written in \texttt{F*} \cite{Swamy:2016}, which is a functional programming language mainly used for program verification. This translation can be used to detect abnormal patterns like \textit{stack overflow} (\ie, exceeding the stack limit). Meanwhile, \cite{kalra2018zeus} proposed Zeus to verify the correctness of smart contracts. Zeus firstly translates the contracts and policy specification into low-level intermediate representation and feeds the encoded representation into constrained horn clauses \cite{mcmillan2007interpolants} to ascertain the safety of the smart contract.

\item \textit{Machine learning based analysis}:
Recently, machine learning-based methods have been proposed to obtain a better representation for detecting vulnerabilities in smart contracts. In particular, \cite{liu2018s} proposed a novel semantic-aware security auditing technique called the $S$-gram scheme for Ethereum. The $S$-gram scheme that combines the $N$-gram language modeling and static semantic labeling can be used to predict potential vulnerabilities by identifying irregular token sequences and optimize existing in-depth analyzers. Meanwhile, the work of \cite{huang2018hunting} translates the bytecode of smart contract into RGB color \HL{that was transformed into images. The images were fed} into a convolutional neural network (CNN) to extract more meaningful features. Moreover, \cite{tann2018towards} applied Long Short Term Memory (LSTM) to analyze the security threats of smart contracts at an opcode level.
\end{itemize}

\subsubsection{Dynamic Control flow}

Despite the fact that the deployed smart contracts are immutable, the control flow of smart contracts is not guaranteed to be immutable. In particular, a smart contract can interact with other contracts (\eg, transferring funds to the contract or creating a new contract). The control flow of smart contract needs to be designed carefully when developing the contract. The interaction of smart contracts can result in an increased number of {interconnected} contracts over time. Therefore, how to predict the contract behaviours becomes challenging. In addition, most of \HL{existing} methods pay attention to \HL{the detection of} potential dynamic control flow problems in programs \HL{while the reliability of the} execution environment is not always ensured. Therefore, it is also significant to check \HL{whether} the execution environment \HL{is reliable}.

\vspace*{0.1cm}
\hspace*{-0.25cm}\textbf{Recent advances {for dynamic control flow}}
\vspace*{0.1cm}

\begin{itemize}
\item \textit{Graph based analysis}: Charlier \etal~\cite{charlierprofiling} proposed a multi-dimensional approach to predict interactions among smart contracts. In particular, this approach integrates stochastic processes and tensors to reproduce existing interactions, consequently predicting future contract interactions. Furthermore, the work in \cite{frowis2017code} presents a heuristic indicator of control flow immutability. In particular, this approach was evaluated on a call graph of all smart contracts on Ethereum. Through analyzing the call graph, it is shown that two smart contracts (out of five) require \HL{a} trust in at least one third party.

\item \textit{Path-searching}: Nikoli{\'c} \etal~\cite{nikolic2018finding} proposed a method namely MAIAN to detect vulnerabilities across a long sequence of invocations of a contract. MAIAN employs inter-procedural symbolic analysis and concrete validator for exhibiting real exploits. It searches the spaces of all execution paths in a trace with depth-first search (DFS) and checks whether the contract triggers property violation. Different from above graph-based methods, MAIAN is designed to identify either locking funds indefinitely, leakage to arbitrary users or being killed by anyone. Therefore, it does not need to model the interactions among smart contracts.

\item \textit{Execution environment}: EVMFuzz~\cite{fu2019evmfuzz} was proposed to detect vulnerabilities of the execution environment of smart contract. EVMFuzz continuously generate seed contracts for different EVM executions, so as to find as many inconsistencies among execution results as possible. This method can eventually discover vulnerabilities with \HL{cross-referencing outputs}.
\end{itemize}

\subsection{Execution challenges}

Execution phase is crucial to smart contracts as it determines the final state of smart contracts. There are a number of issues to be addressed during the execution of smart contracts.

%For example, a smart contract may needs to be executed dynamically (which means the outcome is not deterministic) and the randomness may cause subsequent dispute.

\subsubsection{Trustworthy oracle}
Smart contracts cannot work without real-world information. For example, an Eurobet (\ie, a soccer betting smart contract) needs to know the result of European Cup. However, a smart contract is designed to run in a sandbox isolating from the outside network. In a smart contract, an \emph{oracle} plays a role of an agent who finds and verifies real-world occurrences and forwards this information to the smart contract. Thus, how to determine a trustworthy \textit{oracle} becomes a challenge.

\vspace*{0.1cm}
\hspace*{-0.25cm}\textbf{Recent advances {for trustworthy oracle}}
\vspace*{0.1cm}

\begin{itemize}

\item \textit{Third-party involved}: Town Crier (TC) \cite{zhang2016town} was proposed to address this challenge. In particular, TC scrapes data from reliable web sites and feeds those data to smart contracts. TC feeds the data in the form of datagram that is accompanied with the specific data-source web site and a concrete time frame. Meanwhile, TC executes its core functionality in a Software Guard Extension (SGE) enclave that protects TC from \HL{attacks of} malicious behaviors. %Since TC codes will also be published, results from TC are reliable.

\item \textit{Decentralized}: Ref. \cite{adler2018astraea} proposed a decentralized oracle named ASTRAEA, which is based on a voting game among stake-holders. \HL{In particular, v}oters place a reasonable amount of stakes to vote the random proposition selected from the system. Once the weighted sum of votes matches the vote from a voter, the voter will be rewarded, otherwise, the voter will be penalized. Meanwhile, \cite{ellis2017decentralized} proposed a smart contract based solution for selecting trustworthy oracles. A reputation contract is used to record each oracle-service-provider's reputation according to its previous performance. Then an aggregating contract will calculate the final results of a query from users and finalize the result.
\end{itemize}

\subsubsection{Transaction-ordering dependence}

Users send transactions to invoke functions in a smart contract while miners pack the transactions into blocks. However, the order of transactions is not deterministic due to the uncertainty of the \HL{bisected} blockchain branches \cite{zibin2016blockchain}. This uncertainty can cause inconsistency of order-dependent transactions. For example, there is a contract containing variant $x$. Alice sends a transaction to increase $x$ by 1 while Bob sends a transaction to multiply $x$ by 10. Due to \HL{uncertainty} of the transaction order, the final outcomes on variant $x$ can either be $x+1$ or $x\times 10$. It is worth mentioning that this inconsistency has been well solved in conventional database management systems (DBMS) \cite{Silberschatz:db2010} while it is challenging to solve it in smart contracts as far as we know.

\vspace*{0.1cm}
\hspace*{-0.25cm}\textbf{Recent advances {for transaction-ordering dependence}}
\vspace*{0.1cm}

%The work of \cite{luu2016making} proposes a solution to the transaction-ordering dependence. In particular, this method considers a rule to standardize the execution of {smart contract}  codes to fulfill the specified order requirement. For example, the execution of the code can either turn out to be the expected result or the failure. As a result, the transaction-ordering dependence problem can be solved.
\begin{itemize}

\item \textit{Sequential execution}: Ref. \cite{mavridou2017designing} introduced a design pattern of smart contract-transaction counter. Transaction counter expects a transition number in each function as a parameter and ensures the number be increased by one after each function execution. Through \HL{analyzing} the transition number, the inconsistency problem is solved.

\item \textit{Predefining contract}: To avoid such anomaly, \cite{natoli2016blockchain} proposed to write smart contracts instead of transactions. For example, if Alice wants to increase the value of $x$ after Bob's operation, she \HL{writes} a \texttt{IncreaseIfMultiplied()} function, which avoids the situation where Alice's operation executes prior to Bob's.

\end{itemize}

\subsubsection{Execution Efficiency}

Smart contracts are serially executed by miners. In other words, a miner will not execute another contract until the current contract is completed. The execution serialization essentially limits the system performance. However, it is challenging to execute smart contracts concurrently due to the shared data between multiple smart contracts. In the meantime, how to inspect the contract data without prescribed interface is also important to improving the smart contract execution efficiency as it removes the need to redeploy a new contract.

\vspace*{0.1cm}
\hspace*{-0.25cm}\textbf{Recent advances {for execution efficiency}}
\vspace*{0.1cm}

\begin{itemize}
\item \textit{Execution serialization}: To fill this gap, Dickerson \etal~\cite{dickerson2017adding} proposed an approach based on Software Transactional Memory to allow miners or validators to execute contracts in parallel. The main idea of this approach is to treat each invocation of a smart contract as a speculative atomic action. In this way, conflicts happened during the parallel executions can be rolled back easily. Furthermore, the work in \cite{sergey2017concurrent} investigated smart contracts in a concurrent perspective. In particular, concurrency issues such as atomicity, interference, synchronization, and resource ownership have been well studied in this paper. Ref. \cite{anjana2018efficient} proposed to use optimistic Software Transactional Memory Systems to help improve the execution efficiency of smart contracts. While executing contract transactions concurrently using multi-threading, the miner also stores a block graph of transactions into the block. Then the validators re-execute the smart contract concurrently with the given block graph. If the result is consistent, the block will be appended into the blockchain.

\item \textit{Inspection of contract}: After deployment, the contract content cannot be modified. What can developers do if they are asked to observe some values that are not described in their initial requirements? A straightforward solution \HL{is to} amend the smart contract and re-deploy it. However, the redeployment of smart contracts may cause additional costs. Ref. \cite{bragagnolo2018smartinspect} proposed to exploit the memory layout \emph{reification} to decompile the binary structure of a compiled contract. Meanwhile, the work of \cite{bragagnolo2018smartinspect} proposed the decompilation capabilities encapsulated in mirrors \cite{bracha2004mirrors}, through which the method can introspect the current state of a smart contract instance without redeploying it.
\end{itemize}
%
%and proposed the analogy: \textit{Accounts using smart contracts in a blockchain are like threads using concurrent objects in shared memory}. They compare the similarities and differences between smart contract and concurrent objects. For example,
%smart contracts also have internal mutable states and can be accessed by multiple parties as concurrent objects do. But in contrast to concurrent objects, smart contracts methods are atomic due to the transaction model.
%
%In this paper, we explore remarkable similarities between multi-transactional behaviors of smart contracts in cryptocurrencies such as Ethereum and classical problems of shared-memory concurrency. We examine two real-world examples from the Ethereum blockchain and analyzing how they are vulnerable to bugs that are closely reminiscent to those that often occur in traditional concurrent programs. We then elaborate on the relation between observable contract behaviors and well-studied concurrency topics, such as atomicity, interference, synchronization, and resource ownership. The described contracts-as-concurrent-objects analogy provides deeper understanding of potential threats for smart contracts, indicate better engineering practices, and enable applications of existing state-of-the-art formal verification techniques.

\subsection{Completion challenges}

%Despite the fact that the number of smart contract applications are increasing rapidly, the smart contract technology still face privacy problems. Additionally, the readability of the contract will be a great hindrance to the development of smart contract technology if it is not well addressed.

After the execution of smart contract, the modification to the states in the system will be packed as a transaction and broadcasted to each node. However, the proliferation of smart contracts brings additional concerns.

\subsubsection{Privacy and Security}
Most current smart contract and blockchain platforms lack of privacy-preserving mechanisms, especially for transactional privacy. In particular, the transaction records (\ie, the sequence of operations) are disseminated throughout the whole blockchain networks. Consequently, all the transactions are visible to everyone in the networks. Although some blockchain systems utilize pseudonymous public keys to improve the anonymity of the transactions, most transaction data (such as balances) are still publicly visible. As shown in \cite{Ron2013}, it is possible to obtain useful information from the transaction data based on the transactional graph analysis. {smart contract}  systems also have \HL{their} inherent software vulnerabilities, which are susceptible to malicious attacks. In addition, smart contracts run on top of blockchain systems which are also suffering \HL{from} system vulnerability. For example, it is reported in \cite{apostolaki2017hijacking} that attackers exploited Border Gateway Protocol (BGP) routing scheme to intercept messages in blockchains. It can cause high delay of message broadcasting and also hijack the traffic of a subset of nodes, thereby stealing digital currency.

\vspace*{0.1cm}
\hspace*{-0.25cm}\textbf{Recent advances {for privacy and security}}
\vspace*{0.1cm}

\begin{itemize}

\item \textit{Privacy}: To address the privacy concerns of smart contracts, Kosba \etal~\cite{kosba2016hawk} proposed Hawk - a decentralized smart contract system to establish \HL{privacy-preserved} smart contracts. In particular, the Hawk compiler will compile a contract into a cryptographic protocol automatically. The compiled Hawk program contains two major parts: a private portion used to execute the major function and a public portion used to protect users. Hawk will encrypt the transaction information (\eg, transaction balance) and verify the correctness of transactions via using zero-knowledge proofs (\ie,  without viewing the content of the transactions). The anonymity of the parties in smart contracts can be ensured while the secrecy of contract execution may not be fulfilled. Enigma \cite{zyskind2015enigma} offers a solution to the secrecy of smart contract execution. Advanced cartographic algorithms are used in Enigma to support zero-knowledge proofs. Moreover, Enigma \HL{distributes} blockchain data in different nodes unlike traditional blockchain redundant schemes (\ie, every node saves a copy of all transactions).

\item \textit{Security}: There are some efforts in solving the security concerns. For example, the recent work of \cite{Apostolaki:2018} proposes a secure relaying-network for blockchains, namely SABRE. In particular, SABRE adjusts the inter-domain routing policies for BGP routing scheme. It can protect the link between clients and relays via placing relays appropriately. Meanwhile, SABRE also adopts the co-design of hardware and software via software defined networking (SDN) to reduce the traffic burden at relays. Experimental results demonstrate the effectiveness against BGP routing attacks.	
\end{itemize}

\begin{figure}
\begin{mylisting}[hbox,label=code,drop shadow]
contract Multiplicator {
    ...
    function multiplicate(address receiver) payable {

    if (msg.value >= this.balance)
     {receiver.transfer(this.balance+msg.value);}

}
\end{mylisting}
\caption{Example of smart contract honeypot \cite{torres2019art}}
\label{fig:honeypot}
\end{figure}

\subsubsection{Scams}
As a new technology, blockchain and smart contracts are vulnerable to \HL{malicious attacks initiated} by scams. \HL{The detection of scams is of great importance especially for contract users since it enables them to terminate their investments at an early phase to avoid the unnecessary loss}.

\vspace*{0.1cm}
\hspace*{-0.25cm}\textbf{Recent advances {for readability challenge}}
\vspace*{0.1cm}

\begin{itemize}

\item \textit{Ponzi scheme}: Ponzi scheme is a classical fraud which promises high \HL{return rates} with little risk to investors. It pays the older investors with new investors' funds. But if there is no enough \HL{circulating} money, the scheme \HL{unravels} those posteriors who consequently lose their money. The recent work of \cite{bartoletti2017dissecting} conducted a systematic study over the Ponzi schemes on Ethereum. In particular, 16,082,269 transactions \HL{were} collected from July, 2015 to May, 2017. It was found that 17,777 transactions were related to Ponzi schemes, which \HL{had already collected} over 410,000 US dollars within only two years. Chen \etal~\cite{chen2018detecting} proposed \HL{a method to extract features from both accounts and the operation codes to identify Ponzi schemes on Ethereum}. Meanwhile, the work of \cite{bartoletti2018data} proposed a novel approach to detect and quantify Ponzi schemes \HL{on Bitcoin}. In particular, to address the difficulty \HL{of identifying} Ponzi schemes \HL{as they} often use multiple addresses, a clustering method \cite{bartoletti2018data} was proposed to identify the addresses. They found that 19 out of 32 Ponzi schemes use more than one addresses.

\item \textit{Honeypot}: Smart contract Honeypot implies that the vulnerable-looking contracts contain hidden traps. Take {Figure}  \ref{fig:honeypot} as an example. At the first glance, a naive user may believe that the contract will automatically return the total amount of current balance plus extra money after he or she sends the money to this contract. However, the balance will increase prior to the function execution and the condition of \texttt{if (msg.value >= this.balance)} will never \HL{be satisfied}. The work of \cite{torres2019art} developed a taxonomy of honeypot techniques and use symbolic execution and heuristics to detect honeypots in smart contracts. In addition, \cite{torres2019art} found that the honeypot contracts have made over \$90,000 profit for \HL{creators}.
\end{itemize}

\section{{Smart contract development }platforms}
\label{platforms}

Recently, smart contracts have been developed on block-chain-based platforms. These platforms provide developers with simple interfaces to build {smart contract}  applications. Among a number of incumbent blockchain platforms, many of them can support smart contracts. In this paper, we introduce 5 most representative {smart contract}  platforms: Ethereum \cite{buterin2013Ethereum}, Hyperledger Fabric \cite{cachin2016architecture}, Corda \cite{brown2016corda}, Stellar \cite{Stellar}, Rootstock \cite{RSK} in Section \ref{subsec:platforms}. We choose them mainly due the popularity in developing community and technical maturity as implied in \cite{Bocek2018}. We next summarize the common features of them in Section \ref{subsec:compare}. Finally, we give an example of developing a smart contract in Section \ref{subsec:example}.

\subsection{Representative platforms}
\label{subsec:platforms}

\subsubsection{Ethereum}
Ethereum is a decentralized platform that can execute smart contracts. In contrast to Bitcoin's \textit{Turing-incomplete} script system, Ethereum has developed \textit{Turing-complete} languages such as Solidity\footnote{https://solidity.readthedocs.io/en/develop/}, Serpent\footnote{https://github.com/ethereum/wiki/wiki/Serpent}, Low-level Lisp-like Language (LLL)\footnote{https://lll-docs.readthedocs.io/en/latest/lll\_introduction.html} and Mutan\footnote{https://github.com/ethereum/go-ethereum/wiki} to support general user applications beyond cyptocurrency applications. 	Ethereum compiles \HL{smart contracts} written by Solidity, Serpent, LLL and Mutan into machine codes, which will then be loaded to EVM and run. Meanwhile, Ethereum adopts the account-based data model, in which each participant is identified by its digital wallet.

Similar to Bitcoin, Ethereum adopts PoW as the consensus algorithm, which is also computational intensive. To compensate the cost of solving puzzles done by miners, \emph{Ether} (ETH) instead of coins (BTC) in Bitcoin is used. Essentially, \emph{gas} serves as an internal price for executing a transaction to overcome the unstable value of ETH. Informally, the total cost of a transaction can be calculated by $\textsf{\small gas\_limit} \times \textsf{\small gas\_price}$, where $\textsf{\small  gas\_limit}$ denotes the maximum amount of gas to be used to generate a block and $\textsf{\small gas\_price}$ is the cost of a unit of gas (in ETH). Users can pay different amounts of gas to let their transactions be confirmed \HL{earlier} or later (\ie, large amount of gas resulting in the fast confirmation). \HL{Since} PoW is \HL{computationally} intensive, it can waste a number of electricity for meaningless block mining \HL{tasks}. \HL{It is expected if the mining process is used for meaningful events, such as help solving mathematical puzzles and conducting machine learning tasks.}

\begin{table*}[t]
\caption{Comparison of Smart Contract Platforms}
\label{tab:platforms}
\renewcommand{\arraystretch}{1.5}
\centering
\footnotesize
\begin{tabular}{p{2.5cm} |p{2.2cm} p{2.2cm} p{2.2cm} p{2.2cm} p{2.2cm}p{2.2cm}}
\hline
&\bf{Ethereum} &\bf{Fabric} &\bf{Corda} & \bf{Stellar} & \bf{Rootstock}&\bf{EOS}\\ \hline\hline
Execution environment & EVM & Docker & JVM & Docker & VM &WebAssembly\\\hline
Language & Solidity, Serpent, LLL, Mutan & Java, Golang & Java, Kotlin & Python, JavaScript, Golang and PHP, etc. & Solidity &C++\\\hline
Turing Completeness & Turing complete & Turing complete & Turing incomplete & Turing incomplete &  Turing complete&  Turing complete \\\hline
Data model & Account-based & Key-value pair & Transaction-based & Account-based & Account-based &Account-based\\\hline
Consensus & PoW & PBFT & Raft & Stellar Consensus Protocol (SCP) & PoW&BFT-DPOS \\\hline
Permission & Public & Private & Private & Consortium & Public &Public\\\hline
Application & General & General & Digital currency & Digital currency & Digital currency&General \\
\hline
\end{tabular}
\end{table*}

\subsubsection{Hyperledger Fabric}
Hyperledger Fabric is also a distributed ledger platform for running smart contracts \cite{cachin2016architecture}. Different from Ethereum who runs smart contacts in virtual machines (\ie, EVM), Hyperledger adopts Docker container to execute the code. In contrast to virtual machines (VMs), containers can support {smart contract}  applications with lower overhead while sacrificing the isolation (\ie, applications in one container are running on top of one operating system). Instead of developing {smart contract}  languages of Etherum, Fabric supports conventional high-level programming languages such as Java and Go ({\it aka} Golang). Similarly, Fabric is also Turing complete. Fabric adopts \emph{key-value} pair as the data model.

As Fabric is designed to support general enterprise applications, the Fabric blockchain-network is permissioned (private or consortium). Users have to be authorized to join the network by certificate authorities (CAs). Since there are different roles in the network, multiple types of CAs coexist. For instance, the \textit{enrollment certificate authority} (ECA)  allows users to register with blockchains. Once the user has registered, he/she has to request transaction certificates from \textit{transaction certificate authority} (TCA). The consensus can be easily reached within the permissioned blockchain-network. Fabric exploits PBFT which requires the multi-round voting among authenticated users. {{PBFT relies on multi-round communications among nodes, which can cause time delay. More efficient consensus algorithms should be developed to resolve this problem.}}

%
%
%ecute the code. Additionally, in contrast to Ethereum's permissionless design (\ie, any one can join Ethereum blockchain network), Hyperledger Fabric is a permissoned blockchain (\ie, one has to be authorized to join the network). \textit{Chaincode} and \textit{Certificate Authority} (CA) are important parts in Fabric. Moreover, many traditional high-level languages including Golang and Java can be used in Fabric while only contract-oriented languages (such as Solidity, Serpent and Mutan) are allowed in Ethereum.
%We next show a smart contract sample running in Fabric.

%\subsubsection{Chaincode}
%Chaincode is a piece of code running on blockchains. Three core interfaces are implemented in chaincode: 1) \textit{Init} function will be called when the chaincode is deployed; 2) \textit{Query} will be called when users want to query chaincode state; 3) \textit{Invoke} function will be called when chaincode functions are called to conduct real transactions.
%
%\subsubsection{CA}
%As Fabric is designed for enterprises, the Fabric blockchain network is permissioned. Users have to be authorized to join the network by certificate authorities (CAs). Since there are different roles in the network, multiple types of CAs coexist. For instance, the \textit{enrollment certificate authority} (ECA)  allows users to register with blockchains. Once the user is registered, he/she has to request transaction certificates from \textit{transaction certificate authority} (TCA).

\subsubsection{Corda}

In contrast to diverse applications of Ethereum, Corda \cite{brown2016corda} is specialized for digital-currency applications. It serves as a distributed-ledger platform for saving and processing historical digital-asset records. Corda adopts high-level programming languages such as Java and Kotlin\footnote{https://kotlinlang.org/}, which are running on top of Java Virtual Machine (JVM). Meanwhile, Corda is Turing incomplete to support the verifiability. Moreover, the data model in Corda is the transaction-based model.

Corda typically supports private platforms, in which enterprises establish an authored network to exchange digital-assets in the private manner. In private blockchain platforms, the consensus can easily reach. Corda adopts Raft \cite{Howard:2015} as the consensus algorithm. The consensus in Raft can be achieved by selecting a leader, log replication and safety assurance. Instead of globally broadcasting in blockchains, Corda uses the point-to-point messaging mechanism. Users have to specify the message receivers and the detailed information to be sent.

\subsubsection{Stellar}

Similar to Corda, Stellar \cite{Stellar} is a specialized platform for digital-currency applications. Compared with Ethereum, Stellar is simpler and more accessible. Meanwhile, Stellar can support a diversity of languages such as Python, JavaScript, Golang and PHP. However, Stellar contracts are not Turing complete. Similar to Fabric, Stellar executes program codes on top of Docker containers, consequently reducing the overhead. For example, the execution cost of one transaction at Stellar is only $\sim$\$0.0000002, which can almost be ignored. Moreover, the execution time for one transaction in Stellar is about 5 seconds on average in contrast to 3.5 minutes in Ethereum. Therefore, Stellar is an ideal platform for digital-currency applications. Like Ethereum, Stellar adopts the account-based model as the data model. Stellar developed its own consensus algorithm - Stellar Consensus Protocol (SCP) \cite{Stellar}. Since Stellar is permissioned, the consensus can be easily reached via SCP.

\subsubsection{Rootstock}

Rootstock (RSK) \cite{RSK} runs on top of Bitcoin while supporting faster execution of transactions. For example, RSK can confirm the executed transaction within 20 seconds. Meanwhile, RSK is compatible with Ethereum (\eg, adopting Solidity to implement contracts). RSK contracts are also Turing complete\footnote{Bitcoin is not Turing complete while Rootstock is Turing complete.}. Furthermore, RSK developed its own virtual machines to run smart contracts. The data model of RSK is also account-based while RSK is a public blockchain system. RSK developed its consensus scheme based on PoW while it adopts lightweight implementation consequently reducing the overhead. Like Corda and Stellar, RSK was proposed to mainly support digital-currency applications. \HL{RSK has a merit, i.e., much safer than those systems independent of blockchains since it runs on top of Bitcoin}. However, it can cause extra burden on Bitcoin blockchain. How to resolve this problem is crucial to RSK.

\subsubsection{EOS}
EOS \cite{io2017eos} is designed to enable the scalability of decentralized applications. Instead of using one type of consensus algorithms \HL{only}, EOS combines Byzantine Fault Tolerance (BFT) and Delegated Proof of Stake (DPOS), \HL{thereby} obtaining the advantages of both consensus algorithms. At each round, delegates (\HL{\ie, block producers}) will be selected by stake holders to produce a new block and BFT will proceed among those selected delegates to make the block irreversible. Similar to Bitcoin, EOS is also account-based but it also allows all accounts to be referenced by human readable names. Instead of customizing a virtual machine for code execution like Ethereum, EOS chooses to use WebAssembly (Wasm) so that it is possible to write a smart contract in various languages as long as it can be compiled into Wasm (\eg, EOS supports C++).

\subsection{Comparison of Smart Contract Platforms}
\label{subsec:compare}
Table \ref{tab:platforms} compares Ethereum, Fabric, Corda, Stella, Rootstock (RSK) and EOS from the following aspects such as execution environment, supporting language, Turing completeness, data model, consensus protocols, permission and application. We next summarize the main characteristics of these representative {smart contract}  platforms as follows.
\begin{itemize}
\item{\it Execution Environment}. Contracts in Ethereum are executed in EVM. Similarly, Corda and Rootstock adopt JVM and RSK virtual machines, respectively. In contrast, Fabric and Stellar run smart contracts on top of Docker containers, consequently reducing the overhead while sacrificing the isolation of applications. EOS chooses to use Wasm to support more smart contract languages.

\item{\it {Supported Languages}}.
Ethereum supports Solidity, Serpent and Mutan, which are specially designed for Ethereum. Fabric currently supports Java and Golang while Corda adopts Java and Kotlin. Stellar can support a diversity of languages such as Python, Javascript, Golang and PHP. To be compatible with Ethereum, RSK adopts Solidity as the contract language \HL{while EOS currently only} supports C++.

\item{\it Turing completeness}. Smart contracts on Ethereum, EOS, Fabric and RSK are all Turing complete while Corda and Stellar are Turing incomplete. Turing-complete contracts are typically more expressive than Turing-incomplete contracts. However, the Turing completeness also brings the potential software bugs \HL{being} susceptible to malicious attacks (as illustrated in Section \ref{pra}).

\item{\it Data Model}.
Corda adopts the unspent transaction output (UTXO) model as Bitcoin does. In UTXO model, each payment has to specify the previous unspent transaction as the input. Then the specified transaction becomes \textit{spent}. The changes will be made on new unspent transactions. Ethereum, Stellar, EOS and RSK adopt account-based model, in which the balance of an address is recorded directly instead of calculating all the unspent transaction amounts. Fabric exploits key-value model, in which data is represented in key-value pairs stored in blockchains.

\begin{figure*}[t]
 \centering
 \includegraphics[width=17.6cm]{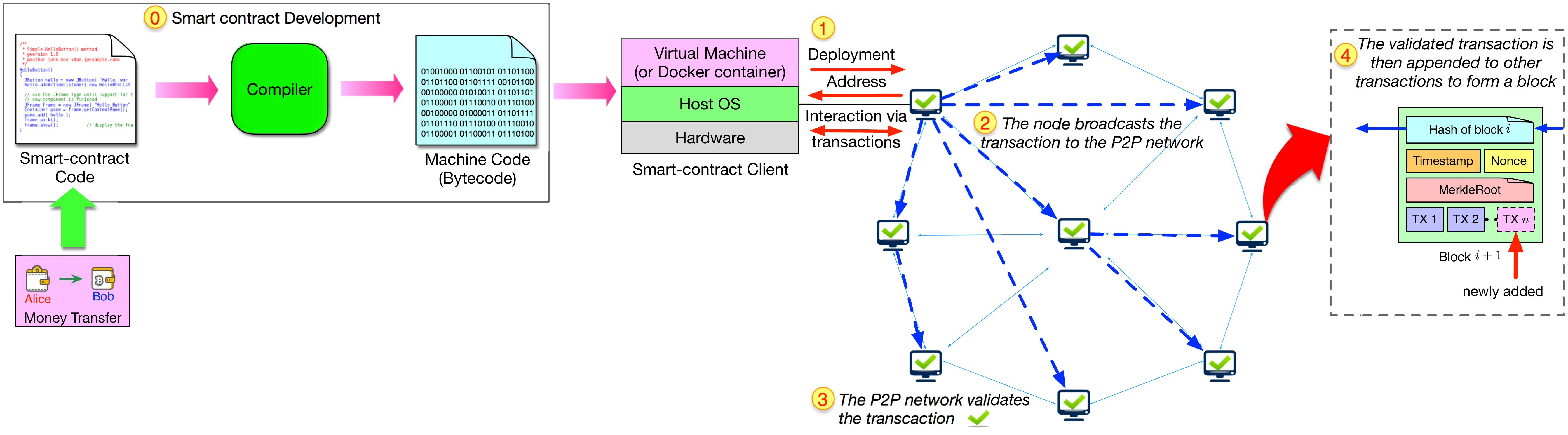}
 \caption{{Workflow} of a smart contract}
 \label{fig:working-contract}
\end{figure*}

\item{\it Consensus Algorithms}.
Ethereum and RSK adopt PoW, in which the validation of the trustfulness of a block is equivalent to the solution of a computationally difficult problem (\ie, a puzzle). PoW consensus algorithms are typically computational-intensive. Fabric chooses to PBFT consensus algorithm \cite{castro1999practical}, in which several rounds of voting among authenticated nodes are taken to reach the consensus. Therefore, PBFT is network-intensive. In contrast, Corda adopts a simple consensus algorithm namely Raft to achieve the consensus between different sectors at the level of individual deals instead of a global system. Similarly, Stellar develops a simply consensus algorithm named SCP to reach the consensus. EOS uses the combination of BFT and DPOS.

\item{\it Permission}.
Ethereum, EOS and RSK are public (\ie permissionless) {smart contract}  platforms and each user can arbitrarily join the network while Corda and Hyperledger are private platforms only allowing authenticated users to access. Stellar sitting between public and private blockchains is a consortium blockchain across different enterprise sectors (or organizations).

\item{\it Applications {of smart contract}}.
Unlike Corda, Stellar and RSK only \HL{support} digital-currency while Ethereum and Fabric cater for a wider diversity of applications ranging from digital currency, digital-asset management, capital investment, public sector to sharing economy. In the future, Corda, EOS, Stellar and RSK and their derivatives may support more general applications.

\end{itemize}

%\bf{Permission} &Public &Permissioned & Permissioned   \\\\
%\bf{Support Languages} &Solidity, Serpent and Mutan &Java, Golang  &Java, Kotlin \ \\\\
%\bf{Consensus} & PoW &PBFT & None \\\\
%\bf{Data Model} &Account-based & Account-based & UTXO-based\\\\
%\bf{Execution Environment} &EVM &Docker&JVM \\\\

\subsection{Example of developing a smart contract}
\label{subsec:example}

\begin{figure}[b]
\begin{mylisting}[label=code]
// mortal.sol
pragma solidity ^0.4.0;
contract mortal {
    /* Define variable owner of the type address */
    address owner;
    /* This function is executed at initialization */
     constructor() internal { owner = msg.sender; }
    /* Function to send 500 wei to receiver's address; 10^18 wei=1 ether; */
    function fundtransfer(address receiver) public { if (msg.sender == owner) {receiver.transfer(500);} }
}
\end{mylisting}
\caption{Example of smart contract written in Solidity in Ethereum}
\label{fig:solidity}
\end{figure}

We next show how to develop and deploy a smart contract. Take a contract of money transfer between Alice and Bob as an example as shown in {Figure}  \ref{fig:working-contract}. After several rounds of negotiations, the agreement between Alice and Bob reaches. Then the agreement is implemented by a {smart contract}  language (\eg, Solidity in Ethereum and Golang in Fabric). The {smart contract}  code is next compiled via a compiler (\eg, solc for Solidity), which generates machine code (or bytecode) running on top of either virtual machines (\eg, EVM, JVM) or Docker containers at a {smart contract}  client. The {smart contract}  client is essentially connected through a peer-to-peer network. After the smart contract is deployed across the blockchain network, a unique contract address is returned to the client to support the future interactions. Thereafter, users can interact with the blockchain network via executing transactions in the smart contract (\eg, deducting the specified amount of money from Alice's digital wallet and increasing the corresponding amount of money in Bob's wallet). It is worth mentioning that each transaction needs to be validated across the blockchain network via the consensus algorithms as shown in {Figure}  \ref{fig:working-contract}. The validated transaction is then appended to the list of transactions. Since every node has a copy of the updated blockchain, it is difficulty to falsify the blockchain data.

\emph{Coding Sample.}
The syntax of Solidity is similar to JavaScript and it also supports inheritance and user-defined types. {Figure}  \ref{fig:solidity} shows an example of a smart contract written in Solidity.

\section{Applications { of smart contract}}
\label{application}

\begin{figure*}[t]
 \centering
 \includegraphics[width=13.5cm]{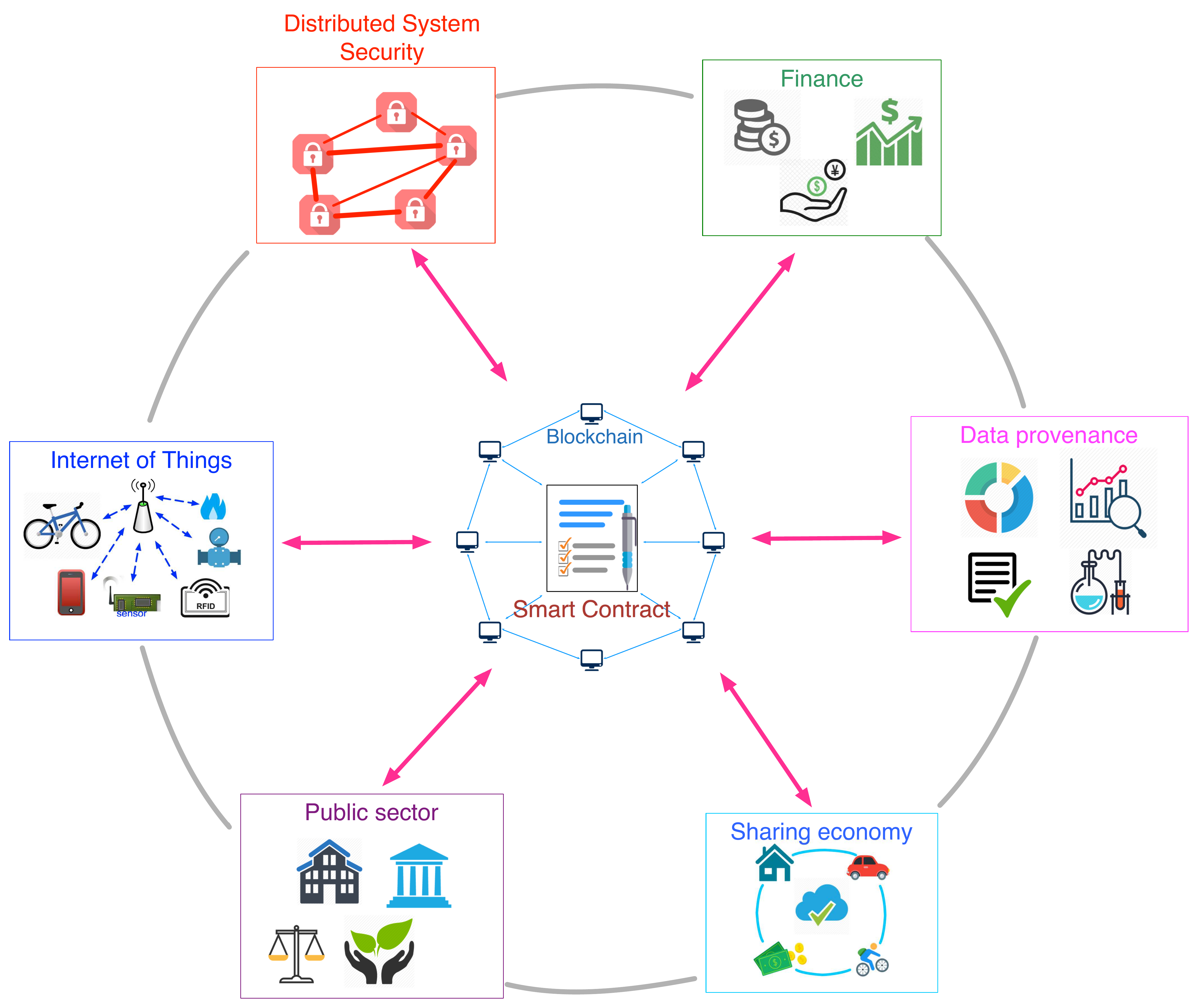}\\
 \caption{Smart contract applications}
 \label{fig:usecases}
\end{figure*}

Smart contracts have a broad spectrum of applications ranging from Internet of Things to sharing economy. In particular, we roughly categorize major smart contract applications into six types as shown in {Figure}  \ref{fig:usecases}. We next describe them in details.
%Smart contract is proven to has the potential to revolutionize our lives with so many smart contract applications springing up in diverge domains. In this section, we will present five typical smart contract application domains: \textit{Internet of Things, Security, Society, Distributed systems} and \textit{Sharing economy}. {Figure}   \ref{fig:usecases} shows example of smart contract applications.

 \subsection{Internet of Things}
Internet of things (IoT) that is one of the most promising technologies can support numerous applications including supply chain management, inventory control systems, retailers, access control, libraries, e-health systems, industrial Internet \cite{habeeb2018real,khattak2019perception,yaqoob2017rise}. The main initiative of IoT is to integrate ``\textit{smart}'' objects (\ie, \HL{``things''}) into the Internet and to provide various services to users \cite{Miorandi:AdHocNets2012}. IoT has been proposed to automate various business transactions in an implicit way.

With the integration with smart contracts, the potentials of IoT can be unleashed. Take industrial manufacturing as an example. Most current manufacturers maintain their IoT ecosystems in a centralized manner. For instance, firmware updates can only be obtained at the central server \emph{manually} by various IoT devices through querying from devices to the server. Smart contracts offer an automatic solution to this problem \cite{christidis2016blockchains}. Manufacturers can place firmware update hashes on smart contracts deployed on blockchains distributed throughout the whole network. Devices can then obtain the firmware hashes from smart contracts automatically. In this way, resources are greatly saved. 

Smart contracts can also bring benefits to IoT e-business model. For example, the traditional e-business model often requires a third party serving as an agent to complete the payment. However, this centralized payment is costly and cannot fully utilize advantages of IoT. In \cite{zhang2015iot}, Distributed autonomous Corporations (DACs) \HL{were} proposed to automate transactions, in which there are no traditional roles like governments or companies involved with the payment. Being implemented by smart contracts, DACs can work automatically without human intervention. Moreover, smart contracts can also help to speed up conventional supply chains. For example, the marriage of supply chains with smart contracts can automate contractual rights and obligations during the payment and the delivery of goods while all the parties in the whole process are trustful.

\subsection{Distributed system security}
Smart contracts can bring benefits in improving the security of distributed systems. Distributed Denial-of-Service (DDoS) attacks are one of major security threats in computer networks. Attackers flood the targeted machine with superfluous requests to overload systems, consequently interrupting or suspending Internet services~\cite{mansfield2015growth}. Recently, a collaborative mechanism was proposed to mitigate DDoS attacks \cite{rodrigues2017blockchain}. Compared with traditional solutions, this scheme \HL{that is based on smart contracts can} tackle the attacks in a fully decentralized manner. In particular, once a server is attacked, the IP addresses of attackers will be automatically stored in a smart contract. In this manner, other nodes will be informed of the addresses of attackers. Furthermore, other security polices will be immediately enforced, \eg, filtering the traffic from the malicious users.

Cloud computing is a promising technology to offer a ubiquitous access of a shared pool of computing and storage resources to users \cite{HWang:ICDCS17}. Generally, users can purchase cloud services from trustful cloud service providers (CSPs). However, how to verify the trustfulness of CSPs becomes a challenge since CSPs often collude with each other to earn more profits. Dong \etal~\cite{dong2017betrayal} proposed a solution based on game theory and smart contracts. The main idea of this approach is to let a client ask two cloud servers to compute the same task. During this process, smart contracts are used to stimulate tension, betrayal and distrust between the clouds. In this way, users can easily determine the rational clouds that will not collude and cheat. Experiments based on the contracts written in Solidity on the official Ethereum network were also conducted to verify the effectiveness of this proposal.

Moreover, brokers are typically used in cloud computing. Users' requests are checked by a broker to match with providers' services. However, both users and service providers must trust the broker. Once the broker is hijacked or compromised, both the parties become untrustful. Recently, Scoca \etal~\cite{scoca2017smart} proposed to use smart contracts to avoid the usage of brokers. The main idea of their approach is to use distributed Service-Level-Agreements for Clouds (dSLAC) \cite{uriarte2016dynamic} to specify the needs via smart contracts. Meanwhile, a utility function that evaluates the agreements according to both parties' preferences was proposed to solve the mismatching problem.

\subsection{Finance}

Smart contracts can potentially reduce financial risks, cut down administration and service costs and improve the efficiency of financial services. We next explain the benefits of smart contracts in the following typical financial services.

\begin{itemize}
\item \textit{Capital markets and investment banking}. The traditional capital markets have suffered from the long settlement cycles. Smart contracts can significantly shorten the settlement period from 20 days or more to 6 to 10 days consequently increasing attractiveness to customers. As a result, it is predicted in \cite{Capgemini:2016} that it can bring 5\% to 6\% demand growth in the future and lead to additional income.

\item \textit{Commercial and retail banking}. In additional to capital markets, the adoption of smart contracts can also bring benefits to the mortgage loan industry \cite{guo2016blockchain}. Conventional mortgage loans are typically complicated in the origination, funding and servicing processes, consequently causing extra costs and delays. Smart contracts can potentially reduce the costs and the delays through automating the mortgage processes with the digitization of legal documents in blockchains.

\item \textit{Insurance}. The application of smart contracts in the insurance industry can also reduce the processing overheads and save the costs especially in \HL{claim handling} \cite{tapscott2017blockchain}. Take the motor insurance as an example. There are multiple parties in a motor insurance: insurer, customers, garages, transport providers and hospitals \cite{du2001secure}. Smart contracts can automate the settlement of claims by sharing legal documents in the distributed ledger consequently improving the efficiency, reducing the claim processing time and saving costs. {For another example, the insurance gaint AXA has launched its insurance for flight delay based on Ethereum smart contracts. Passengers who purchase flight insurances will automatically sign a smart contract, which connects to the global air traffic database. If the system notices a flight delay of over two hours, it will trigger a function in the smart contract, thereby passengers being paid immediately}.
\end{itemize}

\subsection{Data provenance}

In addition to financial services, smart contracts can also be used to ensure information quality in scientific research and public health.  It is reported in \cite{george2016research} that fabrication or falsification of data in clinical trials have occurred frequently in recent years. For another example, one paper published in Nature in 2009 was reported to contain fraud data conducted by Haruko Obokata \cite{rasko2015pushes}. The fabricated data can either mislead the ongoing research directions or hamper the recovery of patients. Consequently, it can seriously undermine the scientific and public trust.

Data provenance has been subsequently proposed to mitigate this problem. The main idea of data provenance is to store the meta-data information of data origin, derivation and transformation. However, there are a number of challenges in enforcing data provenance. For example, most provenance logging tools such as Progger \cite{ko2014progger} and Trusted Platform Module (TPM) \cite{Taha:trustcom2015} store data activities along with privacy-sensitive information (\eg, user ID, accessing time and user roles). How to preserve the privacy information is a challenge. Ramachandran and Kantarcioglu \cite{ramachandran2017using} proposed a data provenance system based on smart contracts and blockchains. Researchers can submit their encrypted data to this system. When there are any data changes, smart contracts will be invoked to track the transformations made to the data. In this manner, any malicious falsifications of data can be captured.

In addition, Liang \etal~\cite{Liang:CCGRID2017} proposed ProvChain to collect and verify cloud data provenance. The main idea of ProvChain is to embed provenance data into blockchain transactions so that any data modifications are traceable. ProvChain consists \HL{of} three procedures: provenance data collection, provenance data storage and provenance data validation. Experimental results demonstrate that ProvChain offers tamper-proof data provenance, privacy-preservation and data reliability.

\begin{figure}[t]
 \centering
 \includegraphics[width=8.5cm]{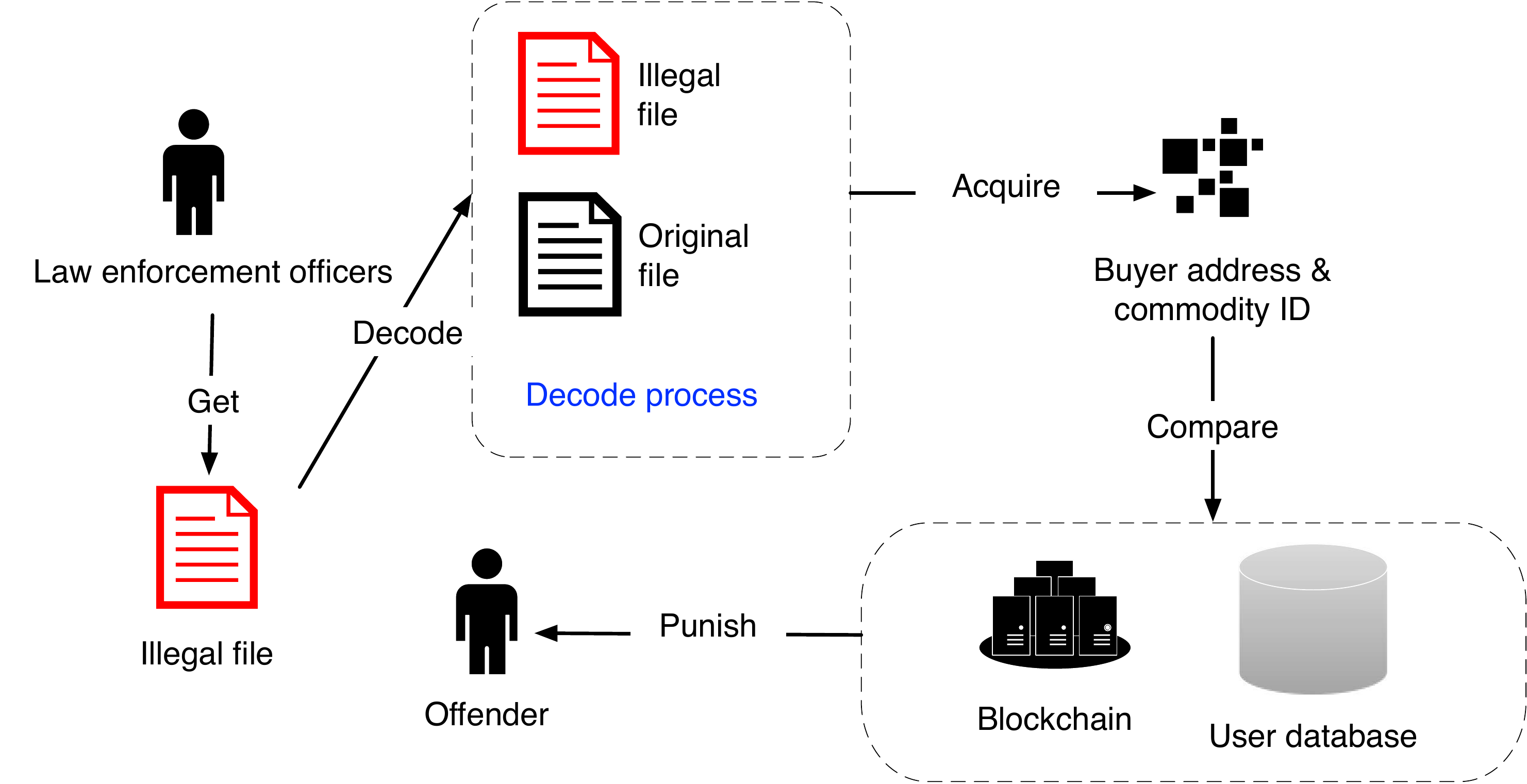}\\
 \caption{Smart contract to protect intellectual property}
 \label{fig:int_prop}
\end{figure}

 {Moreover, smart contract can be used to protect \emph{intellectual property} of creative digital media. For example, Figure \ref{fig:int_prop} shows an example of this application. Each digital product is embedded with a unique digital watermark (such as buyer's digital wallet address and product ID). If there is any infringement (e.g., the buyer sells the digital product to others without the permission from the creator), the law enforcement officer can trace the illegal file with the original file via extracting the digital watermark and comparing the digital wallet address with buyer's. As a result, the infringement of property-right can be easily identified. The whole procedure can be achieved through smart contracts and blockchains.}

\begin{table*}[t]
\centering
\caption{Comparison of smart contract applications}
\label{tab:usecases}
\renewcommand{\arraystretch}{1.5}
\begin{tabular}{m{3.5cm}|m{7cm}|m{6cm}}
\hline
\textbf{Application} & \textbf{Benefits} & \textbf{Use Cases}\\
\hline\hline
\multirow{3}{3.5cm}{Internet of Things \cite{christidis2016blockchains,zhang2015iot}} & \checkmark  Reducing the cost for maintaining central server & (1) IoT device firmware auto-updating\\
&  \checkmark  Automating P2P business trading & (2) Supply chains speeding up\\
& \checkmark  Reducing cost for trusted third parties & \\
\hline

\multirow{3}{3.5cm}{Distributed Systems Security \cite{rodrigues2017blockchain,dong2017betrayal,scoca2017smart}} &  \checkmark  Sharing attack list quickly and reliably & (1) Mitigating DDoS attack in computer networks\\
&  \checkmark Verifying the trustfulness of cloud service providers & (2) Cloud computing\\
& \checkmark Avoiding usage of brokers & \\
\hline

\multirow{3}{3.5cm}{Finance \cite{Capgemini:2016,guo2016blockchain,tapscott2017blockchain}} & \checkmark Reducing financial risks & (1) Capital markets and investment banking\\
& \checkmark Lowering administration and service costs & (2) Commercial and retail banking\\
&\checkmark Improving efficiency of financial services & (3) Insurance\\
\hline

\multirow{3}{3.5cm}{Data Provenance \cite{ramachandran2017using,Liang:CCGRID2017}} &\checkmark Capturing malicious data falsification & (1) Scientific research\\
& \checkmark Improving data reliability & (2) Public health\\
& \checkmark Preserving privacy& (3) Cloud data provenance\\
\hline

\multirow{3}{3.5cm}{Sharing Economy \cite{bogner2016decentralised,huckle2016internet,Xu:2017}} & \checkmark Reducing consumer costs & (1) Item sharing \\
& \checkmark  Reducing cost for trusted third parties  & (2) P2P automatic payment systems \\
& \checkmark  Preserving privacy  & (3) Currency exchange platforms\\
\hline

\multirow{3}{3.5cm}{Public sector \cite{mccorry2017smart,yasin2016online,hillbom2016applications}}& \checkmark Preventing data fraudulence & (1) E-voting systems \\
& \checkmark Data transparency of public information & (2) Personal reputation systems\\
& \checkmark  Preserving privacy  & (3) Smart property exchange platforms\\
\hline
\end{tabular}
\end{table*}%

\subsection{Sharing economy}

The sharing economy brings many benefits such as reducing consumer costs by borrowing and recycling items, improving resource usage, enhancing quality of service, lowering the environment impacts \cite{Taeihagh:2017}. However, most current sharing economy platforms are suffering from high transaction costs of customers, privacy exposure and unreliability of trusted third parties due to the centralization. Smart contracts can potentially reshape sharing economy by decentralizing sharing economy platforms.

Bogner \etal~\cite{bogner2016decentralised} proposed a novel sharing economy platform based on Ethereum smart contracts. In particular, this system allows users to register and share their items without a trusted third party. Meanwhile, personal information is also privacy-preserved. The practical implementation also verifies the effectiveness of the system. In addition, the fusion of Internet of Things (IoT) and smart contracts can also \HL{advance} sharing economy applications. Huckle \etal~\cite{huckle2016internet} discussed the integration of IoT with blockchains to develop sharing economy applications such as peer-to-peer automatic payment systems, traveling systems,  digital assets management and currency exchange platforms.

Meanwhile, a privacy respecting approach was proposed in \cite{Xu:2017} for blockchain-based sharing economy applications. This scheme mainly solves the privacy leakage problem of blockchain-based systems due to the public openness of blockchains. In particular, a zero-knowledge approach was applied to this system. Realistic implementation also demonstrates the effectiveness of the proposed mechanism.

\subsection{Public sector}

Smart contracts along with blockchain technology are also reshaping the public sector management. Blockchain can essentially prevent data fraudulence and provide the transparency of public information. Take a public bidding as an example. The integration of blockchains and smart contracts can prove identities of both bidders and bidding entities, automate the bidding process, provide auditing and reviewing supports.

There are several challenges in e-voting systems, such as user identity verification and user privacy preservation (or voting anonymity). Smart contracts also offer the solution to e-voting systems. A blockchain-based voting system named \HL{\emph{Follow My Vote}}\footnote{Follow My Vote \url{https://followmyvote.com/}} was proposed to verify user identities without the disclosure of user privacy. However, it still relies on a trusted third authority to shuffle the voters so as not to reveal user privacy. McCorry \etal~\cite{mccorry2017smart} utilized the knowledge of {self-tallying} voting protocols (\ie, voters can count the votes without a trusted third party) to build a fair voting system based on smart contracts. In this way, votes can be kept privately while user identities are verifiable at the same time.

Smart contracts can also be used to establish personal digital identity and reputation. For example, Tsinghua University User Reputation System (TURS) \cite{yasin2016online} is an online identity management system based on smart contracts. The TURS profile of a person is based on three aspects: personal reputation, online reputation and professional reputation. Users can protect their private information via smart contracts that grant access permissions to other users by programmable clauses (statements). Meanwhile, all the transactions that are recorded into blockchains cannot be tampered with or removed.

Hillbom and Tillstr{\"o}m \cite{hillbom2016applications} proposed a smart property ownership exchange protocol based on the smart property concept firstly proposed by Szabo \cite{szabo1997formalizing}. In this protocol, each party in the transaction communicates with each other via Bitmessage \cite{warren2012bitmessage}. After the negotiation of the trading details (\eg, a car's digital certificate issued by its manufacturer), the buyer constructs and signs a raw transaction that reassigns the property ownership to the buyer himself/herself. After the signed transaction is sent to the seller, the seller then checks the transaction information. If it is correct, the seller signs on the received transaction and broadcasts it publicly. Moreover, to ensure the consistency, the whole ownership transfer process has to be conducted in an atomic way. In other words, any failure during the whole process will abort the whole ownership transfer process. For example, if the seller does not sign the transaction, he/she will not get the funds from the buyer. Moreover, Li \etal~\cite{Li:TII2017} proposed a secure energy trading system based on consortium blockchain technology. In particular, a credit-based payment scheme was proposed to support fast energy trading without a trusted intermediary.

\textit{Summary}. Table \ref{tab:usecases} compares the smart contract applications. As shown in Table \ref{tab:usecases}, smart contracts can bring numerous benefits for the aforementioned applications. In summary, smart contracts have the merits like reducing the dependence on the trusted third parties, lowering the cost, improving the data reliability and offering privacy-preservation.

\section{Conclusion}
\label{conclusion}

This article presents an overview on the state-of-the-art of smart contracts. In particular, we first provide a brief review on smart contract and blockchain technologies. We then point out the challenges in smart contracts in different aspects of creation, deployment, execution, completion \HL{of smart contracts}. Meanwhile, we also discuss the recent advances in solving these \HL{challenges}. We next compare several major smart contract platforms. Moreover, we categorize smart contract applications and enumerate several typical use cases in each category of applications. In summary, we hope this paper will serve as a guidance to developing secure and scalable \HL{smart contract} applications and promote the evolvement of blockchain technologies.

{On top of blockchains, smart contract is developing rapidly albeit there are still a number of challenges to be addressed.  Most of current research topics on smart contract focus on programming language, security and privacy issues while the proliferation of blockchain and smart contract applications also \HL{poses} new challenges. Like \HL{other computer software tools}, smart contracts \HL{also} contain a number of bugs which are generated unintentionally or mischievously. However,  \HL{detecting and identifying these bugs} will require extensive efforts in aspects of software engineering and data analytics. In addition, \HL{although} enterprise practitioners lack of knowledge in computing programming, they have the expertise in operational technology and law making (i.e., making contracts), which \HL{however} is the deficiency of computer programmers. How to fill the gap between operational technology (OT) and information technology (IT) is of great importance to the development of smart contracts. The integration of software technology, natural language processing and artificial intelligence is a possible remedy to this challenge in the future. }

\section*{Acknowledgement}
The work described in this paper was supported by the National Key Research and Development Program (2016-YFB1000101), the National Natural Science Foundation of China under (61472338) and the Fundamental Research Funds for the Central Universities. Imran's work is supported by the Deanship of Scientific Research at King Saud University through the research group project number RG-1435-051. The authors would like to thank anonymous reviewers who have provided constructive comments greatly improving the paper.

%%%%%%%%%%%%%%%%%%%%%%%%%%%%%%%%%%%%%%%%%%%%%%%%%%%%%%%%%%%%%%%%%%%%%%%%%%%%%%%%%%%%%

%\bibliography{ijmso}
%\bibliographystyle{unsrt}
%\bibliographystyle{alpha}
%\bibliography{reference}
%\bibliographystyle{./IEEEtran}
%\bibliography{./IEEEabrv,./reference}

%\bibliographystyle{elsarticle-num}
\bibliographystyle{unsrt}
\bibliography{reference}

%\EOD
\end{document}